\documentclass[10pt,a4paper,final]{iopart}
\expandafter\let\csname equation*\endcsname\relax
\expandafter\let\csname endequation*\endcsname\relax
\usepackage[margin=1 in]{geometry}
\usepackage{amsmath,color,iopams,graphicx}
\usepackage[breaklinks=true,colorlinks=true,linkcolor=blue,urlcolor=blue,citecolor=blue]{hyperref}
\usepackage[english]{babel}
\usepackage{graphicx,comment}
\usepackage{mathtools}
\usepackage{amsmath}
\usepackage{footnote}
\makesavenoteenv{tabular}
\makesavenoteenv{table}

\newcommand{\shuru}{\begin{equation} \begin{aligned}}
\newcommand{\khatam}{ \end{aligned}\end{equation}}
\newcommand{\shuruq}{\begin{equation*} \begin{aligned}}
\newcommand{\khatamq}{ \end{aligned}\end{equation*}}

\begin{document}

\title{Density-Density Correlation Function of  Strongly Inhomogeneous Luttinger Liquids}
\author{  Nikhil Danny Babu, Joy Prakash Das, Girish S. Setlur$^{*}$}
\address{Department of Physics \\ Indian Institute of Technology  Guwahati \\ Guwahati, Assam 781039, India}
\ead{$^{*}$gsetlur@iitg.ernet.in}

\begin{abstract}
In this work, we show in pedagogical detail that the {\it{most singular}} contributions to the slow part of the asymptotic density-density correlation function of Luttinger liquids with fermions interacting mutually with only short-range forward scattering and also with localised scalar static impurities (where backward scattering takes place) has a compact analytical expression in terms of simple functions that have second order poles and involve only the scale-independent bare transmission and reflection coefficients. This proof uses conventional fermionic perturbation theory resummed to all orders, together with the idea that for such systems, the (connected) moments of the density operator all vanish beyond the second order - the odd ones vanish identically and the higher order even moments are less singular than the second order moment which is the only one included. This important result is the crucial input  to the recently introduced ``Non-Chiral Bosonization Technique" (NCBT) to study such systems. The results of NCBT {\it{cannot}} be easily compared with the results obtained using conventional bosonization as the former only extracts the most singular parts of the correlation functions albeit for arbitrary impurity strengths and mutual interactions. The latter, ambitiously attempts to study all the parts of the asymptotic correlation functions and is thereby unable to find simple analytical expressions and is forced to operate in the vicinity of the homogeneous system or the half line (the opposite extreme). For a fully homogeneous system or its antithesis viz. the half-line, all the higher order connected moments of the density vanish identically which means the results of chiral bosonization and NCBT ought to be the same and indeed they are.
\end{abstract}

\pacs{71.10.Pm,  73.21.Hb, 11.15.Tk}

\vspace{2pc}
\noindent{\it Keywords}: Luttinger liquid, Correlation Functions,Bosonization


\section{Introduction}

\noindent The presence of impurities in one dimensional systems continues to be a challenging problem in quantum many body physics
\cite{kane1992transport,artemenko2005low,dinh2010tunneling,kainaris2018transmission,lo2019crossover,kane1992transmission,kane1992resonant}. A good number of analytical \cite{grishin2004functional,matveev1993tunneling,kane1992transport,kane1992transmission,samokhin1998lifetime}  and numerical approaches \cite{qin1996impurity,hamamoto2008numerical,freyn2011numerical,ejima2009luttinger,moon1993resonant} have made their mark in the long history of strongly correlated 1D systems. Kane and Fisher in their seminal works \cite{kane1992transport,kane1992transmission} have shown how the nature of the electron-electron interactions( i.e repulsive or attractive) affect the transport properties across barriers or constrictions in a single-channel Luttinger liquid. They elucidated that for repulsive interactions the electrons are always reflected by a weak link and for attractive interactions the electrons are transmitted even through a strong barrier. In \cite{kane1992resonant} they studied the effect of electron interactions on resonant tunneling in presence of a double barrier and showed that the resonances are of non-Lorentzian line shapes with a width that vanishes as $T\rightarrow 0$, in  striking contrast to the noninteracting one dimensional electron gas. However,  analytical expressions of the correlation functions of such systems with arbitrary strength of mutual interactions and in presence of a scatterer of arbitrary strength are yet to be a part of the literature.
Nevertheless, the attempts to reach this central goal of calculating the most general result has led to many peripheral results where the arbitrariness of one or more parameters had to be compromised. The most prevalent analytical tool to deal with these systems, which belong to the universal class of Luttinger liquids \cite{haldane1981luttinger}, is bosonization where a fermion field operator is expressed as the exponential of a bosonic field \cite{giamarchi2004quantum,von1998bosonization}. However the results of this method are exact only for the extreme cases of very weak impurity and very strong impurity (apart from some isolated examples such as the Bethe ansatz solutions \cite{rylands2016quantum,haldane1981demonstration}) and to deal with the general result one has to rely on perturbative approach in terms of the impurity strength (or inter-chain hopping in the other extreme) and renormalize the series to obtain a finite answer \cite{kane1992transport}.\\

\noindent
A recently developed alternative to this, which goes by the name `Non chiral bosonization technique (NCBT)', does a better job of avoiding RG methods and tackling impurities of arbitrary strengths \cite{das2018quantum,das2019nonchiral}. But it can yield only the most singular part of the Green functions. 
 In a recent work \cite{10.1088/1402-4896/ab957f} the four-point Green functions in the context of Friedel oscillations in a Luttinger Liquid were calculated using NCBT and the most singular contribution to the slow part of the local density oscillations were obtained in the form of power laws. Closed analytical expressions for the dynamical density of states exponents were also obtained.
  On the other hand, Matveev et al. \cite{matveev1993tunneling} dealt with impurities of arbitrary strengths using fermionic renormalization only and without using bosonization methods, however their results are valid only for weak strengths of mutual interactions between the fermions.
\\

\noindent
The study of the density correlations of a Luttinger liquid is important, which is well reflected in the literature.  Iucci et al. obtained a closed-form analytical expression for the zero-temperature Fourier transform of the $2k_F$ component of the density-density correlation function in a spinful Luttinger liquid  \cite{iucci2007fourier}.
Schulz studied the density correlations in a one dimensional electron gas interacting with long ranged Coulomb forces and calculated the $4k_F$ component of the density which decays extremely slowly and represents a 1D Wigner crystal \cite{schulz1993wigner}.
Parola presented an exact analytical evaluation of the asymptotic spin-spin correlations of the 1D Hubbard model with infinite on-site interaction ($U \to \infty $) and away from half filling \cite{parola1990asymptotic}. Their results suggested that the renormalization-group scaling to the Tomonaga-Luttinger
model is exact in the $U \to \infty $ Hubbard model.
Stephan et al. calculated the dynamical density-density correlation function for the one-dimensional, half-filled Hubbard model extended with nearest-neighbor repulsion for large on-site repulsion compared to hopping amplitudes \cite{stephan1996dynamical}.
Caux et al. studied the dynamical density density correlations in a 1D Bose gas with a delta function interaction using a Bethe-ansatz-based numerical method \cite{caux2006dynamical}.
Gambetta et al. have done a study of the correlation functions in a one-channel finite size Luttinger liquid quantum dot \cite{gambetta2014correlation}.
Protopopov et al. investigated the four-point correlations of a Luttinger liquid in a non-equilibrium setting \cite{protopopov2011many}. In \cite{sen1999density} Sen et al. performed a numerical study of the Luttinger liquid type behaviour of the density-density correlation functions in the lattice Calogero-Sutherland model. Aristov analyzed the modified density-density correlations when curvature in the fermionic dispersion is present \cite{aristov2007luttinger}.
\\ \mbox{  }\\
 In addition to this, numerical methods such as the density matrix renormalization group (DMRG) \cite{PhysRevLett.69.2863,RevModPhys.77.259} have been employed to study gapped fermionic and spin systems in 1D \cite{Schneider2006ConductanceIS,PhysRevLett.113.070601,Ejima_2009,PhysRevB.56.9766}. However the system which we are interested in viz. Luttinger Liquids with impurities, are gapless and it is difficult to justify the application of DMRG to such systems. But notable attempts have been made in this direction \cite{PhysRevB.99.121103,PhysRevLett.116.247204,PhysRevLett.118.017201,Oshi}. The main achievement of the NCBT method is the careful redefinition of the meaning of the random phase approximation (RPA) in the context of strongly inhomogeneous Luttinger Liquids which is by no means an obvious extension of the corresponding notion in homogeneous systems. This redefinition involves a systematic truncation and resummation of the perturbation series in powers of the fermion-fermion coupling with the impurities being arbitrary. However, DMRG methods (and conventional bosonization) apply it to the full system without any such truncation carried out. As a result we don't expect favourable comparisons between DMRG and NCBT except in limiting cases (half line and fully homogeneous system).
 The truncation and resummation method employed in our work is not a shortcoming since it is only when such a procedure is implemented, closed analytical expressions for the N-point functions of the system become feasible. There is no other limit in which it is possible to write down closed formulas in terms of elementary functions of positions and times for the N-point functions of a system as complicated as mutually interacting fermions in presence of impurities.
\\ \mbox{  }\\
Moreover in DMRG, both forward scattering and backward scattering interactions between the fermions are taken into account. But the model we study using NCBT, includes only short-range forward scattering interactions between the fermions. So our results are bound to be different than what is obtained using techniques like DMRG. We make this distinction clear by taking the example of Oshikawa et.al \cite{Oshi}. Equations (1) and (3) of \cite{Oshi} involve terms such as $ V S_{i}^{z} S_{i+1}^{z}$ . According to Jordan Wigner transformation, $ S^z_i  = n_i - \frac{1}{2} $, and $ n_i = c^{\dagger}_ic_i $ is the full density which includes both the slowly varying and rapidly varying parts. By contrast, our model described in Section 2 considers only the slow part of the density (Eq. \ref{FS} of the present work). Hence a comparison between such approaches and ours is not possible. The reason why we don't include backward scattering between fermions is - the correlation functions cannot be expressed in terms of elementary functions  thereby limiting its use to theoretical physics.
   Also it is unlikely that such numerical methods  are able to capture the most singular parts of the Green’s functions that NCBT provides. The NCBT results  have already been validated analytically using the Schwinger-Dyson equation \cite{das2019nonchiral} and a host of other ways as shown in our group's published works \cite{10.1088/1402-4896/ab957f,das2018quantum,das2017one,das2018ponderous,DAS201939,
Das2019C}.
\\ \mbox{  }\\
In the following sections, it is shown that the most singular contributions (precise meaning defined later) of the slow part of the density density correlation functions of a spinful Luttinger liquid (also spinless) with short-range forward scattering mutual interactions in presence of localized static scalar impurities is shown to be expressible in terms of elementary functions of positions and times which involve only second order poles and also involve only the bare reflection and transmission coefficients of a single particle in the presence of these impurities.

\section{The Model }

\noindent Consider a quantum system that comprises a 1D gas of electrons with forward scattering short-range mutual interactions and in the presence of a scalar potential $V (x)$ that is localized near an origin. The full generic-Hamiltonian of the system contains three parts and can be written as follows.
\shuru
H = H_0+H_{imp} + H_{fs}
\label{Hamiltonian}
\khatam
where $ H_0 $ is the Hamiltonian of {\it free fermions} and $H_{imp}$ is that of the {\it impurity} (or impurities). This has an asymptotic form in terms of its Green function.
$H_{fs}$ is that of {\it short-range forward scattering} mutual interactions between the fermions. 
Using the linear dispersion relations near the Fermi level ($E = E_F + k v_F$) one can write,
\shuru
H_0 =  -iv_F \mbox{ } \int dx \mbox{ }\sum_{\sigma = \uparrow,\downarrow}
 (:\psi^{\dagger}_R(x,\sigma)\partial_x \psi_R(x,\sigma): - :\psi^{\dagger}_L(x,\sigma)\partial_x \psi_L(x,\sigma):)
\khatam
and the impurity Hamiltonian is given to be in Hermitian form as follows (the expression below taken at face value is ill-defined and a regularization procedure is implied as discussed below). \small
\shuru
H_{imp} = &V_0 \mbox{ }\sum_{\sigma = \uparrow,\downarrow} ( \psi^{\dagger}_R(0,\sigma)  \psi_R(0,\sigma) + \psi^{\dagger}_L(0,\sigma) \psi_L(0,\sigma) )
   + V_1 \sum_{\sigma = \uparrow,\downarrow}\psi^{\dagger}_R(0,\sigma) \psi_L(0,\sigma)  + V^{*}_1\sum_{\sigma = \uparrow,\downarrow} \psi^{\dagger}_L(0,\sigma) \psi_R(0,\sigma)
\label{imp}
 \khatam
\normalsize
 where the subscripts R ($\nu = +1$) and L ($\nu = -1 $) are the usual right and left movers. The impurity hamiltonian in equation (\ref{imp}) is ambiguous without proper regularisation. The point of view taken here is that the meaning of equation (\ref{imp}) is indirectly fixed by demanding that the Green function of  $ H_0+H_{imp} $ be given by equation (\ref{twopoint}).
 The Green function of this Hamiltonian $  H_0 + H_{imp} $ may be written as,
\shuru
 <T\mbox{ } \psi_{\nu}(x,\sigma,t)  \psi^{\dagger}_{\nu^{'}}(x^{'},\sigma^{'},t^{'}) >_0  \mbox{} \equiv  \mbox{} \delta_{\sigma,\sigma^{'}}\mbox{ }
\sum_{\gamma,\gamma^{'} = \pm 1 } \mbox{}   G_{\gamma,\gamma^{'}}^{\nu,\nu^{'}}(x,t;x^{'},t^{'})\mbox{  }\theta(\gamma x) \theta(\gamma^{'} x^{'})
\label{twopoint}
\khatam where $ \theta(
x > 0) = 1 $, $ \theta(x < 0) = 0 $ and $ \theta(0) = \frac{1}{2} $ is Heaviside's step function. It can be shown that
 \shuru
  G_{\gamma,\gamma^{'}}^{\nu,\nu^{'}}(x,t;x^{'},t^{'}) = \frac{ g_{\gamma,\gamma^{'}} (\nu,\nu^{'}) }{ \nu x - \nu^{'} x^{'} - v_F(t-t^{'}) }
\khatam
The complex numbers $   g_{\gamma,\gamma^{'}} (\nu,\nu^{'})  $ may be related  to $ V_0$ and $V_1 $, alternatively, to the (bare) transmission ($T$) and reflection  ($R$)  amplitudes.
 In terms of the reflection   and transmission   amplitudes, we have
\shuru
g_{\gamma_1,\gamma_2} (\nu_1,\nu_2)=\frac{i}{2\pi} \left[ \delta_{\nu_1,\nu_2} \delta_{\gamma_1,\gamma_2} +(T \delta_{\nu_1,\nu_2}+R \delta_{\nu_1,-\nu_2})\delta_{\gamma_1,\nu_1}\delta_{\gamma_2,-\nu_2}
+(T^{*} \delta_{\nu_1,\nu_2}+R^{*} \delta_{\nu_1,-\nu_2})\delta_{\gamma_1,-\nu_1}\delta_{\gamma_2,\nu_2}\right]
\khatam
These amplitudes may also be related to the details of the impurity potentials.
The relation between  $ V_0,V_1 $ in equation (\ref{imp}) to the bare transmission and reflection amplitudes is given by (see \hyperref[AppendixE]{Appendix E} for derivation),
\shuru
& V_0 = \frac{2 i {\text{ $v_F$ }} (T-T^*)}{2 T T^*+T+T^*} \\
& V_1 = V_1^{*} =  -\frac{4 i  {\text{ $v_F$ }}\mbox{     } R^* T}{2 T T^*+T+T^*} =
 \frac{4 i {\text{ $v_F$ }}\mbox{     } R T^* }{2 T T^*+T+T^*}
\khatam
\noindent The slow part of the asymptotic density-density correlation may be written down using Wick's theorem as (after subtracting the uncorrelated average product: $\tilde{\rho_s} = \rho_s - <\rho_s>$),
 \shuru
  \\& <T\mbox{   } \tilde{\rho}_s(x,\sigma,t) \tilde{\rho}_s(x^{'},\sigma^{'},t^{'})>_0 \mbox{} =\mbox{}  - \mbox{} \delta_{\sigma,\sigma^{'}}\mbox{ } \sum_{\gamma,\gamma^{'} = \pm 1} \sum_{\nu,\nu^{'} = \pm 1}
  \frac{ |g_{\gamma,\gamma^{'}} (\nu,\nu^{'})|^2 \mbox{  }\theta(\gamma x) \theta(\gamma^{'} x^{'}) }{ (\nu x - \nu^{'} x^{'} - v_F(t-t^{'}))^2 }
\\& =
\mbox{} - \frac{ \delta_{\sigma,\sigma^{'}} }{(2\pi)^2} \mbox{ } (
\mbox{    } \text{sgn}(x_1) \text{sgn}(x_2)  \frac{ |R|^2 }{ ( v_F(t_1-t_2) +  | x_1|+|x_2 |  )^2 }
   + \text{sgn}(x_1) \text{sgn}(x_2)   \frac{ |R|^2 }{  ( v_F (t_1-t_2) - | x_1| - |x_2 |   )^2 }
\\&
+    \frac{1}{ ( v_F (t_1-t_2) +   x_1-x_2    )^2 }
   +   \frac{1}{  (v_F (t_1-t_2) -   x_1+ x_2   )^2 }
  )
  \label{RHORHO0}
\khatam
where,
\shuru
   \rho_s(x,\sigma,t)  =  ( :\psi^{\dagger}_R(x,\sigma,t) \psi_R(x,\sigma,t) : + :\psi^{\dagger}_L(x,\sigma,t) \psi_L(x,\sigma,t) :)
   \khatam
   In the  presence of short-range forward scattering given by the additional piece,
\shuru
H_{fs} = \frac{1}{2}\sum_{\sigma,\sigma^{'} = \uparrow,\downarrow}
\int dx \mbox{ } \int  dx^{'} \mbox{ } v(x-x^{'}) \mbox{ } \rho_s(x,\sigma,t) \rho_s(x^{'},\sigma^{'},t)
\label{FS}
\khatam
it is not possible to write down a simple formula such as equation (\ref{RHORHO0}) for the correlation function described there.  The reason is because $ |g_{\gamma,\gamma^{'}} (\nu,\nu^{'})|^2 $ involves only the bare reflection and transmission coefficients -  but in conventional chiral bosonization, they are renormalized to become scale-dependent. Also there is no guarantee that the function will continue to have simple second order poles as shown in  equation (\ref{RHORHO0}). The {\it{main claim of the non-chiral bosonization technique (NCBT)}} is that if one is willing to be content at the most singular part of this correlation function then it is indeed possible to write down a simple formula very similar to  equation (\ref{RHORHO0}) even when short-range forward scattering i.e. equation (\ref{FS}) is present. Furthermore, this most singular contribution will only involve the bare transmission and reflection coefficients as is the case in equation (\ref{RHORHO0}). This most singular part of the slowly varying asymptotic density-density correlation (ie. DDCF  in presence of equation (\ref{FS})) is  given below and the proof is given  in the next section.
 \shuru
\langle T\mbox{   } \rho_s(x_1,\sigma_1,t_1)\rho_s(x_2,\sigma_2,t_2)\rangle =
\frac{1}{4}\left(\langle T\mbox{   } \rho_h(x_1,t_1)\rho_h(x_2,t_2)\rangle  + \sigma_1\sigma_2
\langle T\mbox{   } \rho_n(x_1,t_1)\rho_n(x_2,t_2)\rangle \right)
\label{Res1}
 \khatam
 where,
\shuru
\langle T\mbox{   } \rho_h(x_1,t_1)\rho_h(x_2,t_2)\rangle  = &\frac{v_F  }{ 2\pi^2 v_h } \mbox{   } \sum_{  \nu = \pm 1 }\bigg ( -  \frac{1}{ ( x_1-x_2 + \nu v_h(t_1-t_2) )^2 }
	-  \frac{\frac{v_F }{v_h}  \mbox{    } \text{sgn}(x_1) \text{sgn}(x_2)\mbox{   }\frac{ |R|^2 }{    \left( 1 -  \frac{(v_h-v_F)}{ v_h } |R|^2   \right) } }{  ( | x_1|+|x_2 | + \nu v_h(t_1-t_2) )^2 }
\bigg)\\
\langle T\mbox{   } \rho_n(x_1,t_1)\rho_n(x_2,t_2)\rangle  = &\frac{1  }{ 2\pi^2  } \mbox{   } \sum_{  \nu = \pm 1 }\bigg ( -  \frac{1}{ ( x_1-x_2 + \nu v_F(t_1-t_2) )^2 }
	-  \frac{ \mbox{    } \text{sgn}(x_1) \text{sgn}(x_2)\mbox{   } |R|^2 }{  ( | x_1|+|x_2 | + \nu v_F(t_1-t_2) )^2 }
\bigg)\\
\label{Res2}
\khatam
with $ v_h = \sqrt{ v_F^2 + \frac{2v_0 v_F}{\pi}} $ and $ \sigma = \uparrow (+1) $ and $ \sigma = \downarrow (-1) $.
\\ \mbox{   } \\
One of the main results of NCBT is the assertion that the most singular contribution to $ <T\mbox{   } \tilde{\rho}_s(x_1,\sigma_1,t_1) \tilde{\rho}_s(x_2,\sigma_2,t_2)> $ in the presence of short-range forward scattering between fermions viz. equation (\ref{FS}) is given by equation (\ref{Res1}) and equation (\ref{Res2}).
\section{Results: Density density correlation function}

\noindent The density density correlation functions (DDCF) in absence of mutual interactions is given by equation (\ref{RHORHO0}). This has to be systematically transformed to include mutual interactions. Firstly, the space time DDCF is related to the momentum frequency DDCF as follows ($ x_1 \neq x_2 $ and $ x_1 \neq 0 $ and $ x_2 \neq 0 $).
\shuru
<T \mbox{   }\rho_s(x_1,t_1;\sigma_1)\rho_s(x_{2},t_{2};\sigma_2)>_0 =
\frac{1}{L^2} \sum_{q,q^{'},n }e^{ - i q x_1 } e^{ - i q^{'} x_2 } e^{ - w_n (t_1-t_2)}
<T \mbox{   }\rho_s(q,n;\sigma_1)\rho_s(q^{'},-n;\sigma_2)>_0
\label{TRNS}
\khatam
From this we can obtain the DDCF in momentum and frequency space as follows (here $\beta$ is the inverse temperature which comes into the calculation because of converting summation to integration which is allowed in the zero temperature limit: $\sum\limits_n f(z_n)=\frac{\beta}{2 \pi} \int f(z) dz $ where $z_n=\frac{ \pi (2n+1)}{\beta}$).
\shuru
\mbox{    }<T \mbox{   }\rho_s(q,n;\sigma_1)\rho_s(q^{'},-n;\sigma_2)>_0\mbox{ }
=\mbox{ }& \frac{ \delta_{ \sigma_1, \sigma_2 }}{ \beta}
 \frac{ (2 v_F q^{'}) (2 v_F q)   |g_{1,1}(1,-1)|^2   |w_n|   (2\pi) }{ ( ( v_F q )^2 +  w^2_n )  ( ( v_F q^{'})^2  +  w^2_n ) }
 \\
 +&  \frac{ \delta_{ \sigma_1, \sigma_2 }}{ \beta}  \frac{2 q^2 v_F}{ w^2_n  + (q  v_F)^2  }    \delta_{ q+q^{'}, 0  }
\frac{ L }{ (2\pi) }
\label{DDCFmf}
\khatam
In \hyperref[AppendixA]{Appendix A}  we show how to recover equation (\ref{RHORHO0}) from equation (\ref{TRNS}) and equation (\ref{DDCFmf}).
Now the generating function for an auxiliary field U in presence of mutual interactions between particles given by $v(x_1-x_2)$ can be written as
\shuru
Z[U] = \int D[\rho ] e^{ i S_{eff,0}[\rho] } e^{ \sum_{q,n,\sigma} \rho_{q,n,\sigma} U_{q,n,\sigma} }
 e^{ -i \int^{-i\beta}_{0} dt \int dx_1 \int dx_2 \frac{1}{2} v(x_1-x_2) \rho(x_1,t_1;.) \rho(x_2,t_1;.)   }
\khatam
where $S_0$ is the action of free fermions and
\shuruq
&\rho(x_1,t_1;.) = \frac{1}{L} \sum_{q,n} e^{ - i q x } e^{ w_n t } \mbox{  }\rho_{q,n;.}\\
&v(x_1-x_2) = \frac{1}{L} \sum_{Q} e^{ -i Q (x_1-x_2) } v_{Q}\\
&\rho(x_1,t_1;.) =  \rho(x_1,t_1;\uparrow) +  \rho(x_1,t_1; \downarrow)\\
\khatamq
Thus the generating function can be written as follows.
\shuru
Z[U] = \int D[\rho ] e^{ i S_{eff,0}[\rho] } e^{ \sum_{q,n,\sigma} \rho_{q,n,\sigma} U_{q,n,\sigma} }
 e^{ - \sum_{q,n}   \frac{\beta v_0}{2L}
\rho_{q,n;.}\rho_{-q,-n;.}
    }
\label{GF1}
\khatam
If one denotes the generating function in absence of interactions as $Z_0$, then
\shuruq
&Z_0[U] = \int D[\rho ] e^{ i S_{eff,0}[\rho] } e^{ \sum_{q,n,\sigma} \rho_{q,n,\sigma} U_{q,n,\sigma} }\\
 \Rightarrow &  e^{ i S_{eff,0}[\rho] }=\int D[U^{'}] e^{ -\sum_{q,n,\sigma} \rho_{q,n,\sigma} U^{'}_{q,n,\sigma} }\mbox{  }Z_0[U^{'}]
\khatamq
Inserting in equation (\ref{GF1}),
\shuru
Z[U] = \int D[\rho ] \mbox{  }\int D[U^{'}] \mbox{  }Z_0[U^{'}]
 \mbox{  } e^{ \sum_{q,n,\sigma} \rho_{q,n,\sigma} (U_{q,n,\sigma}-U^{'}_{q,n,\sigma}) }
 e^{ - \sum_{q,n}   \frac{\beta v_0}{2L}
\rho_{q,n;.}\rho_{-q,-n;.}
    }
\khatam
Set,
\shuru
& \rho_{q,n,\sigma} = \frac{1}{2} \rho_{q,n;.} + \frac{1}{2}\sigma \mbox{  }\sigma_{q,n}
\mbox{ };\mbox{ } U_{q,n,\sigma} = \frac{1}{2} U_{q,n;.} + \frac{1}{2}\sigma \mbox{  }W_{q,n}
\mbox{ };\mbox{ } U^{'}_{q,n,\sigma} = \frac{1}{2} U^{'}_{q,n;.} + \frac{1}{2}\sigma \mbox{  }W^{'}_{q,n} \\\\
\khatam
 Using these relations the generating function can be written as
\shuru
Z[U] = \int D[\rho ] \mbox{  }\int D[U^{'}] \mbox{  }Z_0[U^{'}]
 \mbox{  } &e^{ \frac{1}{4} \sum_{q,n,\sigma} ( \rho_{q,n;.} + \sigma \mbox{  }\sigma_{q,n})
 ( (U_{q,n;.}-U^{'}_{q,n;.}) +\sigma \mbox{  }(W_{q,n}-W^{'}_{q,n}) ) }
 e^{ - \sum_{q,n}   \frac{\beta v_0}{2L}
\rho_{q,n;.}\rho_{-q,-n;.}
    }
\khatam
where in the RPA sense (for the homogeneous system, this choice corresponds to RPA, for the present steeplechase problem this choice corresponds to the most singular truncation of the RPA generating function)
\shuru
Z_0[U^{'}] = e^{ \frac{1}{2} \sum_{q,q^{'},n; \sigma} <\rho_{q,n}\rho_{q^{'},-n}>_0 \mbox{   }U^{'}_{q,n; \sigma}U^{'}_{q^{'},-n;\sigma } }
\label{quadratic}
\khatam
where $<\rho_{q,n}\rho_{q^{'},-n}>_0$ is equation  (\ref{DDCFmf}) with $\sigma_1 = \sigma_2$.   It is to be noted that we have neglected the higher moments of $\rho$ in $Z_0[U']$ beyond the quadratic as they are less singular than the second moment (see section \ref{Gaussian}).  Now,
\shuruq
Z[U] = &\int D[\rho ] \mbox{  }\int D[U^{'}] \mbox{  }e^{ \frac{1}{2} \sum_{q,q^{'},n; \sigma} <\rho_{q,n}\rho_{q^{'},-n}>_0 \mbox{   }\frac{1}{4} (U^{'}_{q,n;.} + \sigma \mbox{  }W^{'}_{q,n})(U^{'}_{q^{'},-n;.} + \sigma \mbox{  }W^{'}_{q^{'},-n}) }
\\
 &\mbox{  } e^{ \frac{1}{4} \sum_{q,n,\sigma} ( \rho_{q,n;.} + \sigma \mbox{  }\sigma_{q,n})
 ( (U_{q,n;.}-U^{'}_{q,n;.}) +\sigma \mbox{  }(W_{q,n}-W^{'}_{q,n}) ) }
 e^{ - \sum_{q,n}   \frac{\beta v_0}{2L}
\rho_{q,n;.}\rho_{-q,-n;.}
    }\\
=& \int D[U^{'}] \mbox{  }e^{ \frac{1}{2} \sum_{q,q^{'},n}
<\rho_{q,n}\rho_{q^{'},-n}>_0 \mbox{   }\frac{1}{2} (U^{'}_{q,n;.}U^{'}_{q^{'},-n;.} + W^{'}_{q,n} W^{'}_{q^{'},-n}) }
\\
& \mbox{  } \int D[\rho ] \mbox{  }e^{ \frac{1}{2} \sum_{q,n} ( \rho_{q,n;.} (U_{q,n;.}-U^{'}_{q,n;.})
  + \sigma_{q,n} (W_{q,n}-W^{'}_{q,n}) ) }
 e^{ - \sum_{q,n}   \frac{\beta v_0}{2L}
\rho_{q,n;.}\rho_{-q,-n;.}
    }\\
= &\int D[U^{'}]\mbox{  }e^{ \frac{1}{4} \sum_{q,q^{'},n}
<\rho_{q,n}\rho_{q^{'},-n}>_0 \mbox{   }  W_{q,n} W_{q^{'},-n} } \mbox{  }
e^{ \frac{1}{4} \sum_{q,q^{'},n}
<\rho_{q,n}\rho_{q^{'},-n}>_0 \mbox{   } U^{'}_{q,n;.}U^{'}_{q^{'},-n;.}  }\\
 &\mbox{  } \int D[\rho ] \mbox{  }e^{ \frac{1}{2} \sum_{q,n}  \rho_{q,n;.} (U_{q,n;.}-U^{'}_{q,n;.})
    }
 e^{ - \sum_{q,n}   \frac{\beta v_0}{2L}
\rho_{q,n;.}\rho_{-q,-n;.}
    }\\
\khatamq
The last result follows from the extremum condition viz.
\shuru
\rho_{-q,-n;.}
 = \frac{L}{2 \beta v_0}  (U_{q,n;.}-U^{'}_{q,n;.})
\khatam
Thus (including only the holon part),
\shuru
Z[U] =& \int D[U^{'}]\mbox{  }
e^{ \frac{1}{4} \sum_{q,q^{'},n}
<\rho_{q,n}\rho_{q^{'},-n}>_0 \mbox{   } U^{'}_{q,n;.}U^{'}_{q^{'},-n;.}  }  \mbox{  }e^{ \frac{1}{4} \sum_{q,n}  \frac{L}{2 \beta v_0}  (U_{-q,-n;.}-U^{'}_{-q,-n;.}) (U_{q,n;.}-U^{'}_{q,n;.})
    }
\khatam
The integration has to be done using the saddle point method. This involves finding the extremum of the log of the integrand with respect to $ U^{'} $ which leads to the answer we are looking for.
This means,
\shuru
 \sum_{q^{'}}
<\rho_{q,n}\rho_{q^{'},-n}>_0 \mbox{   }U^{'}_{q^{'},-n;.}
-  \frac{L}{ 2\beta v_0}  (U_{-q,-n;.}-U^{'}_{-q,-n;.}) = 0
\khatam
Hence,
\shuru
 \beta
\mbox{    }<\rho_{q,n}\rho_{q^{'},-n}>_0
=  \frac{ (2 v_F q^{'}) (2 v_F q)   }{ ( ( v_F q )^2 +  w^2_n )  ( ( v_F q^{'})^2  +  w^2_n ) }
  |g_{1,1}(1,-1)|^2   |w_n|   (2\pi)
 +   \frac{2 q^2 v_F}{ w^2_n  + (q  v_F)^2  }    \delta_{ q+q^{'}, 0  }
\frac{ L }{ (2\pi) }
\khatam
This means,
\shuru
\frac{  (2 v_F q)   }{ ( ( v_F q )^2 +  w^2_n )   }
 &\sum_{q^{'}} \frac{ (2 v_F q^{'})    }{   ( ( v_F q^{'})^2  +  w^2_n ) }
 \mbox{   }U^{'}_{q^{'},-n;.}
  |g_{1,1}(1,-1)|^2   |w_n|   (2\pi)\\
& +   \frac{2 q^2 v_F}{ w^2_n  + (q  v_F)^2  }  \mbox{   }U^{'}_{-q,-n;.}
\frac{ L }{ (2\pi) }
-  \frac{L}{ 2 v_0}  (U_{-q,-n;.}-U^{'}_{-q,-n;.}) = 0
\khatam
Let,
\shuru
\frac{1}{L} \sum_{q^{'}} \frac{ (2 v_F q^{'})    }{   ( ( v_F q^{'})^2  +  w^2_n ) }
 \mbox{   }U^{'}_{q^{'},-n;.} = u^{'}_n
\khatam
or,
\shuruq
\frac{  (2 v_F q)   }{ ( ( v_F q )^2 +  w^2_n )   }
 u^{'}_n L
  |g_{1,1}(1,-1)|^2   |w_n|   (2\pi)
 +   \frac{2 q^2 v_F}{ w^2_n  + (q  v_F)^2  }  \mbox{   }U^{'}_{-q,-n;.}
\frac{ L }{ (2\pi) }
-  \frac{L}{ 2 v_0}  (U_{-q,-n;.}-U^{'}_{-q,-n;.}) = 0
\khatamq
or,
\shuruq
   \left( \frac{L}{ 2 v_0}  +  \frac{ L }{ (2\pi) } \frac{2 q^2 v_F}{ w^2_n  + (q  v_F)^2  }  \right) \mbox{   }U^{'}_{-q,-n;.}
 = -\frac{  (2 v_F q)   }{ ( ( v_F q )^2 +  w^2_n )   }
 u^{'}_n L
  |g_{1,1}(1,-1)|^2   |w_n|   (2\pi)
+  \frac{L}{ 2 v_0}  U_{-q,-n;.}
\khatamq
This means,
\shuru
 U^{'}_{-q,-n;.}
 = -\frac{  (2 v_F q)   }{ ( ( v_F q )^2 +  w^2_n )   }
\frac{  u^{'}_n L
  |g_{1,1}(1,-1)|^2   |w_n|   (2\pi) }{    \left( \frac{L}{ 2 v_0}  +  \frac{ L }{ (2\pi) } \frac{2 q^2 v_F}{ w^2_n  + (q  v_F)^2  }  \right) }
+  \frac{ \frac{L}{ 2 v_0}  U_{-q,-n;.} }{    \left( \frac{L}{ 2 v_0}  +  \frac{ L }{ (2\pi) } \frac{2 q^2 v_F}{ w^2_n  + (q  v_F)^2  }  \right) }
\khatam
Set $ v^2 = v_F^2 + \frac{ 2 v_F v_0 }{\pi} $.
\shuru
 U^{'}_{-q,-n;.}
 = -  (2 v_F q)
\frac{  u^{'}_n 2 v_0
  |g_{1,1}(1,-1)|^2   |w_n|   (2\pi) }{    \left( w^2_n  + (q  v)^2   \right) }
+  \frac{ ( w^2_n  + (q  v_F)^2)  U_{-q,-n;.} }{    \left(  w^2_n  + (q  v)^2   \right) }
\khatam
Hence,
\shuru
u^{'}_n \frac{1}{L} \sum_{q^{'}} \frac{ (2 v_F q^{'})^2    }{   ( ( v_F q^{'})^2  +  w^2_n ) ( w^2_n  + (q^{'}  v)^2 ) }
 \mbox{   }  2 v_0
  |g_{1,1}(1,-1)|^2   |w_n|   (2\pi)
+ \frac{1}{L} \sum_{q^{'}}  \frac{ (2 v_F q^{'})   U_{q^{'},-n;.} }{    \left(  w^2_n  + (q^{'}  v)^2   \right) }
 = u^{'}_n
\khatam
Hence,
\shuru
 u^{'}_n =
 \frac{1}{L} \sum_{q^{'}}  \frac{ (2 v_F q^{'})   U_{q^{'},-n;.} }{    \left( 1 - \frac{4 v_0 v_F}{ (v+v_F) v }
 \mbox{   }
  |g_{1,1}(1,-1)|^2    (2\pi) \right)  \left(  w^2_n  + (q^{'}  v)^2   \right) }
\khatam
or,
\shuru
 \sum_{q^{'}}
<\rho_{q,n}\rho_{q^{'},-n}>_0 \mbox{   }U^{'}_{q^{'},-n;.}
-  \frac{L}{ 2\beta v_0}  (U_{-q,-n;.}-U^{'}_{-q,-n;.}) = 0
\khatam
Hence,
\shuru
Z[U] =
 \mbox{  }e^{ \frac{1}{4} \sum_{q,n}  \frac{L}{2 \beta v_0}  (U_{-q,-n;.}-U^{'}_{-q,-n;.}) U_{q,n;.}    }
\khatam
\shuru
U_{-q,-n;.}- U^{'}_{-q,-n;.}
 =&
  \frac{1}{L} \sum_{q^{'}}  \frac{ (2 v_F q^{'})   U_{q^{'},-n;.} }{    \left( 1 - \frac{4 v_0 v_F}{ (v+v_F) v }
 \mbox{   }
  |g_{1,1}(1,-1)|^2    (2\pi) \right)  \left(  w^2_n  + (q^{'}  v)^2   \right) }
  (2 v_F q)
\frac{ 2 v_0
  |g_{1,1}(1,-1)|^2   |w_n|   (2\pi) }{    \left( w^2_n  + (q  v)^2   \right) }
\\&+ U_{-q,-n;.} \frac{  (q  v)^2  - (q  v_F)^2  }{    w^2_n  + (q  v)^2  }
\khatam
But we had
\shuruq
Z[U] = \int D[U^{'}]\mbox{  }
e^{ \frac{1}{4} \sum_{q,q^{'},n}
<\rho_{q,n}\rho_{q^{'},-n}>_0 \mbox{   } U^{'}_{q,n;.}U^{'}_{q^{'},-n;.}  }
 \mbox{  }e^{ \frac{1}{4} \sum_{q,n}  \frac{L}{2 \beta v_0}  (U_{-q,-n;.}-U^{'}_{-q,-n;.}) (U_{q,n;.}-U^{'}_{q,n;.})
    }
\khatamq
This means
\shuru
\\ Z[U] =
&e^{ \sum_{q,n}  \frac{L}{8 \beta v_0}  U_{q,n;.} U_{-q,-n;.} \frac{  (q  v_h)^2  - (q  v_F)^2  }{    w^2_n  + (q  v_h)^2  }}
\\&\mbox{  }e^{  \sum_{q,n}  \frac{L}{8 \beta v_0}   \frac{1}{L} \sum_{q^{'}}  \frac{ (2 v_F q^{'})    U_{q,n;.} U_{q^{'},-n;.} }{    \left( 1 - \frac{4 v_0 v_F}{ (v_h+v_F) v_h }
 \mbox{   }
  |g_{1,1}(1,-1)|^2    (2\pi) \right)  \left(  w^2_n  + (q^{'}  v_h)^2   \right) }
  (2 v_F q)
\frac{ 2 v_0
  |g_{1,1}(1,-1)|^2   |w_n|   (2\pi) }{    \left( w^2_n  + (q  v_h)^2   \right) } }
\khatam
Here,
\shuru
v_h=\sqrt{v_F^2+\frac{2v_F v_0}{\pi}}
\khatam
Since we have,
\shuru
& \rho_{q,n;.} = \rho_{q,n;\uparrow} + \rho_{q,n;\downarrow}
\mbox{ };\mbox{ } \sigma_{q,n} = \rho_{q,n;\uparrow} - \rho_{q,n;\downarrow}\\
\khatam
Thus,
\shuru
\frac{1}{4} \mbox{  }<\rho_{q,n;.}\rho_{q^{'},-n;.}> &=  \delta_{q+q^{'},0}
   \mbox{  }\frac{L}{4 \beta  }  \frac{  2v_F   q^2  }{   \pi ( w^2_n  + (q  v_h)^2 ) }
\\
+ &  \frac{1}{2\beta} \frac{   (2 v_F q)(2 v_F q^{'}) |g_{1,1}(1,-1)|^2   |w_n|   (2\pi)  }{
   \left( 1 -   \frac{(v_h-v_F)}{  v_h }
 \mbox{   }
  |g_{1,1}(1,-1)|^2    (2\pi)^2 \right)  \left(  w^2_n  + (q^{'}  v_h)^2   \right)   \left( w^2_n  + (q  v_h)^2   \right) }
\mbox{  }
 \\
\khatam

\shuru
\frac{1}{4} \mbox{  }<\sigma_{q,n}\sigma_{q^{'},-n}>&\mbox{ } = \frac{ ( 2v_F q^{'}) (2 v_F q)   }{ 2\beta ( ( v_F q )^2 +  w^2_n )  ( ( v_F q^{'})^2  +  w^2_n ) }
  |g_{1,1}(1,-1)|^2   |w_n|   (2\pi)
 +   \frac{ q^2 v_F L }{ 2 \pi \beta (w^2_n  + (q  v_F)^2)  }    \delta_{ q+q^{'}, 0  }\\\\
<\sigma_{q,n}\rho_{q^{'},-n;.}> \mbox{ }& = \mbox{ }0
\khatam
Finally, the full density density correlation functions in momentum frequency space can be written as,
\shuru
<\rho_{q,n,\sigma}\rho_{q^{'},-n,\sigma^{'}} > = \frac{1}{4} <\rho_{q,n;.}\rho_{q^{'},-n;.} >
 + \frac{1}{4} \sigma \sigma^{'}   <\sigma_{q,n}\sigma_{q^{'},-n} >
\khatam
Doing an inverse Fourier transform of the above we get the DDCF in real space time.
\footnotesize
\shuru
<&T \mbox{   }\rho_s(x_1,t_1;\sigma)\rho_s(x_{2},t_{2};\sigma^{'})> =
\\&
- \frac{v^2_F }{v_h^4}  \mbox{    } \text{sgn}(x_1) \text{sgn}(x_2)
\mbox{    }\left( \frac{ \beta }{2\pi} \mbox{    } \frac{1}{ [ (t_1-t_2) + \frac{| x_1|+|x_2 |  }{v_h}]^2 }
   + \frac{ \beta }{2\pi}  \mbox{    } \frac{1}{  [(t_1-t_2) -\frac{| x_1|+|x_2 | }{v_h} ]^2 } \right)
 \frac{1}{2\beta} \frac{    |g_{1,1}(1,-1)|^2  (2\pi)
  }{    \left( 1 - \frac{4 v_0 v_F}{ (v_h+v_F) v_h }
 \mbox{   }
  |g_{1,1}(1,-1)|^2    (2\pi) \right)    }
\\&
 - \frac{ \text{sgn}(x_1)   \text{sgn}(x_2) }{v_F^2}
 \mbox{  }
\left( \frac{ \beta }{2\pi} \mbox{    } \frac{1}{ [ (t_1-t_2) + \frac{| x_1|+|x_2 |  }{v_F}]^2 }
   + \frac{ \beta }{2\pi}  \mbox{    } \frac{1}{  [(t_1-t_2) -\frac{| x_1|+|x_2 | }{v_F} ]^2 }
  \right)
  \frac{  1  }{ 2\beta  }
  |g_{1,1}(1,-1)|^2    (2\pi  \sigma \sigma^{'} )
\\&
-  \frac{v_F  }{ 4 \pi \beta  v_h^3 } \mbox{   } \left( \frac{ \beta }{2\pi} \mbox{    } \frac{1}{ [ (t_1-t_2) + \frac{| x_1-x_2 |  }{v_h}]^2 }
   + \frac{ \beta }{2\pi}  \mbox{    } \frac{1}{  [(t_1-t_2) -\frac{| x_1-x_2 | }{v_h} ]^2 }
\right)
\\&-   \frac{ \sigma \sigma^{'}  }{ 4 \pi \beta  v_F^2 }
\left( \frac{ \beta }{2\pi} \mbox{    } \frac{1}{ [ (t_1-t_2) + \frac{| x_1-x_2 |  }{v_F}]^2 }
   + \frac{ \beta }{2\pi}  \mbox{    } \frac{1}{  [(t_1-t_2) -\frac{| x_1-x_2 | }{v_F} ]^2 }
 \right)
\label{DENDEN}
\khatam
\normalsize
One can separate the holon and spinon parts of the DDCF and write as follows.
\small
\shuru
\langle T\mbox{   } \rho_h(x_1,t_1)\rho_h(x_2,t_2)\rangle  = &\frac{v_F  }{ 2\pi^2 v_h } \mbox{   } \sum_{  \nu = \pm 1 }\bigg ( -  \frac{1}{ ( x_1-x_2 + \nu v_h(t_1-t_2) )^2 }
	-  \frac{\frac{v_F }{v_h}  \mbox{    } \text{sgn}(x_1) \text{sgn}(x_2)\mbox{   }\frac{ |R|^2 }{    \left( 1 -  \frac{(v_h-v_F)}{ v_h } |R|^2   \right) } }{  ( | x_1|+|x_2 | + \nu v_h(t_1-t_2) )^2 }
\bigg)\\
\langle T\mbox{   } \rho_n(x_1,t_1)\rho_n(x_2,t_2)\rangle  = &\frac{1  }{ 2\pi^2  } \mbox{   } \sum_{  \nu = \pm 1 }\bigg ( -  \frac{1}{ ( x_1-x_2 + \nu v_F(t_1-t_2) )^2 }
	-  \frac{ \mbox{    } \text{sgn}(x_1) \text{sgn}(x_2)\mbox{   } |R|^2 }{  ( | x_1|+|x_2 | + \nu v_F(t_1-t_2) )^2 }
\bigg)\\
\label{full1}
\khatam
\normalsize
The full density density correlation functions can thus be written in a compact form as follows.
\shuru
\langle T\mbox{   } \rho_s(x_1,\sigma_1,t_1)\rho_s(x_2,\sigma_2,t_2)\rangle =
\frac{1}{4}\left(\langle T\mbox{   } \rho_h(x_1,t_1)\rho_h(x_2,t_2)\rangle  + \sigma_1\sigma_2
\langle T\mbox{   } \rho_n(x_1,t_1)\rho_n(x_2,t_2)\rangle \right)
\label{full2}
\khatam

\section{Discussion}

\noindent We have said repeatedly that  equation (\ref{full1}) and equation (\ref{full2}) displays the central result of this work, viz., the most singular part of the density density correlation function of the generic Hamiltonian given in equation  (\ref{Hamiltonian}). In this section, the key aspects of this result are discussed.

\subsection{Gaussian approximation}
\label{Gaussian}

\noindent In equation (\ref{quadratic}), it is seen that only the quadratic moment of $\rho$ in the $Z_0[U']$ is included, neglecting all the higher moments. 	
 The reason for this is the following. The connected parts of the odd moments $ \rho $ vanish identically (in the RPA limit) and that of the even moments are less singular than the second moment, etc. For example, the connected 4-density function,\small
\shuru
< T \rho(x_1) \rho(x_2)\rho(x_3)\rho(x_4) >_c
\mbox{      }    \sim\mbox{      }  &
<T\psi(x_1)\psi^*(x_2)>  <T\psi(x_2)\psi^*(x_3)> <T\psi(x_3)\psi^*(x_4)>  <T\psi(x_4)\psi^*(x_1)> \\
&+\mbox{ }  \text{permutations}\\
\khatam
\normalsize
has only first order poles and no second order poles. The NCBT does not claim to provide the asymptotic Green functions of strongly inhomogeneous interacting Fermi systems exactly but it does claim to provide the {\it{most singular part of}} the asymptotic Green functions of strongly inhomogeneous interacting Fermi systems exactly. However this is not the case for the homogeneous systems ($|R|=0$) and half-lines ($|R|=1$) where both the even and the odd moments higher than the quadratic moment of density functions vanishes. Hence for these extreme cases, the density-density correlation functions are not just the most singular part but the full story. To prove this, a quantity `$\Delta$' is defined as follows.

\shuru
\Delta \equiv& <T\mbox{  } \rho_s(x_1,\sigma_1,t_1)\rho_s(x_2,\sigma_2,t_2)\rho_s(x_3,\sigma_3,\sigma_3,t_3)\rho_s(x_4,\sigma_4,t_4)>_0\\
&-<T\mbox{  } \rho_s(x_1,\sigma_1,t_1)\rho_s(x_2,\sigma_2,t_2)>_0\mbox{   }<T\mbox{ }\rho_s(x_3,\sigma_3,t_3)\rho_s(x_4,\sigma_4,t_4)>_0
\\&
-  <T\mbox{  } \rho_s(x_1,\sigma_1,t_1)\rho_s(x_3,\sigma_3,t_3)>_0\mbox{ } <T\mbox{ }\rho_s(x_2,\sigma_2,t_2)\rho_s(x_4,\sigma_4,t_4)>_0\\
&-
<T\mbox{  } \rho_s(x_1,\sigma_1,t_1)\rho_s(x_4,\sigma_4,t_4)>_0\mbox{ }<T\mbox{ }\rho_s(x_2,\sigma_2,t_2)\rho_s(x_3,\sigma_3,t_3)>_0
\khatam
If $ \Delta = 0 $ it means Wick's theorem is applicable at the level of pairs of fermions which makes the Gaussian theory valid.
So now, the question is whether $\Delta$ is zero or not. Expanding the above expression using conventional Wick's theorem and using the form of the Green function shown in equation (\ref{twopoint}) we have the following results for various cases.\\

\noindent {\it Case I:} All four points on same side ($x_1 > 0, x_2 > 0, x_3 > 0, x_4 > 0 $)\\

\noindent We have
 $ \rho_s(x,\sigma,t) = :  \psi^{\dagger}_R(x,\sigma,t)\psi_R(x,\sigma,t)  +   \psi^{\dagger}_L(x,\sigma,t)\psi_L(x,\sigma,t)  : $ where $::$ represents normal ordering. For this case we obtain,\small
\shuru
<T&\mbox{ } \rho_s(x_1,\sigma,t_1) \rho_s(x_2,\sigma,t_2) \rho_s(x_3,\sigma,t_3) \rho_s(x_4,\sigma,t_4)>_c \mbox{      }  = \mbox{ }
\frac{|R|^2 |T|^2}{2 \pi ^4}\times\\
 &\bigg(\frac{1}{((t_3-t_1) v_F+x_1+x_3) ((t_3-t_2) v_F+x_2+x_3) ((t_4-t_1) v_F+x_1+x_4) ((t_4-t_2) v_F+x_2+x_4)}
\\&
+\frac{1}{((t_1-t_2) v_F+x_1+x_2) ((t_3-t_2) v_F+x_2+x_3) ((t_1-t_4) v_F+x_1+x_4) ((t_3-t_4) v_F+x_3+x_4)}
\\&
+\frac{1}{((t_2-t_1) v_F+x_1+x_2) ((t_3-t_1) v_F+x_1+x_3) ((t_2-t_4) v_F+x_2+x_4) ((t_3-t_4) v_F+x_3+x_4)}
\\&
+\frac{1}{((t_2-t_1) v_F+x_1+x_2) ((t_2-t_3) v_F+x_2+x_3) ((t_4-t_1) v_F+x_1+x_4) ((t_4-t_3) v_F+x_3+x_4)}
\\&
+\frac{1}{((t_1-t_2) v_F+x_1+x_2) ((t_1-t_3) v_F+x_1+x_3) ((t_4-t_2) v_F+x_2+x_4) ((t_4-t_3) v_F+x_3+x_4)}
\\&
+\frac{1}{((t_1-t_3) v_F+x_1+x_3) ((t_2-t_3) v_F+x_2+x_3) ((t_1-t_4) v_F+x_1+x_4) ((t_2-t_4) v_F+x_2+x_4)} \bigg)\\\\
\khatam

\normalsize
\noindent {\it Case II:} Three points on same side ($x_1 < 0, x_2 > 0, x_3 > 0, x_4 > 0 $)\\\small

\shuru
<T&\mbox{ } \rho_s(x_1,\sigma,t_1) \rho_s(x_2,\sigma,t_2) \rho_s(x_3,\sigma,t_3) \rho_s(x_4,\sigma,t_4)>_c \mbox{      }  = \mbox{ }
\frac{|R|^2 |T|^2 }{2 \pi ^4} \times
\\
 \bigg(&-\frac{1}{((t_1-t_3) v_F+x_1-x_3) ((t_3-t_2) v_F+x_2+x_3) ((t_1-t_4) v_F+x_1-x_4) ((t_4-t_2) v_F+x_2+x_4)}
\\&
 -\frac{1}{((t_1-t_2) v_F-x_1+x_2) ((t_3-t_2) v_F+x_2+x_3) ((t_1-t_4) v_F-x_1+x_4) ((t_3-t_4) v_F+x_3+x_4)}
\\&
 -\frac{1}{((t_1-t_2) v_F+x_1-x_2) ((t_1-t_3) v_F+x_1-x_3) ((t_2-t_4) v_F+x_2+x_4) ((t_3-t_4) v_F+x_3+x_4)}
\\&
 -\frac{1}{((t_1-t_2) v_F+x_1-x_2) ((t_2-t_3) v_F+x_2+x_3) ((t_1-t_4) v_F+x_1-x_4) ((t_4-t_3) v_F+x_3+x_4)}
\\&
 -\frac{1}{((t_1-t_2) v_F-x_1+x_2) ((t_1-t_3) v_F-x_1+x_3) ((t_4-t_2) v_F+x_2+x_4) ((t_4-t_3) v_F+x_3+x_4)}
\\&
 +\frac{1}{((t_3-t_1) v_F+x_1-x_3) ((t_2-t_3) v_F+x_2+x_3) ((t_1-t_4) v_F-x_1+x_4) ((t_2-t_4) v_F+x_2+x_4)} \bigg)\\\\
\khatam

\normalsize
\noindent {\it Case III:} Two points on same side ($x_1 < 0, x_2 < 0, x_3 > 0, x_4 > 0 $)\\\small

\shuru
<T&\mbox{ } \rho_s(x_1,\sigma,t_1) \rho_s(x_2,\sigma,t_2) \rho_s(x_3,\sigma,t_3) \rho_s(x_4,\sigma,t_4)>_c \mbox{      }  = \mbox{ }
\frac{|R|^2 |T|^2 }{4 \pi ^4}
\\
\bigg (&-\frac{1}{((t_1-t_3) v_F+x_1-x_3) ((t_3-t_2) v_F-x_2+x_3) ((t_1-t_4) v_F+x_1-x_4) ((t_2-t_4) v_F+x_2-x_4)}
\\&
 -\frac{2}{((t_3-t_2) v_F+x_2-x_3) ((t_1-t_3) v_F-x_1+x_3) ((t_1-t_4) v_F-x_1+x_4) ((t_2-t_4) v_F-x_2+x_4)}
\\&
 -\frac{1}{((t_2-t_3) v_F+x_2-x_3) ((t_3-t_1) v_F-x_1+x_3) ((t_1-t_4) v_F+x_1-x_4) ((t_2-t_4) v_F+x_2-x_4)} \bigg)
\\&
+ \frac{R^{*2} T^2 }{4 \pi ^4} \bigg(-\frac{1}{((t_1-t_2) v_F+x_1+x_2) ((t_1-t_3) v_F+x_1-x_3) ((t_2-t_4) v_F-x_2+x_4) ((t_3-t_4) v_F+x_3+x_4)}
\\&
-\frac{1}{((t_1-t_2) v_F+x_1+x_2) ((t_2-t_3) v_F-x_2+x_3) ((t_1-t_4) v_F+x_1-x_4) ((t_4-t_3) v_F+x_3+x_4)}
\\&
-\frac{1}{((t_2-t_1) v_F+x_1+x_2) ((t_1-t_3) v_F-x_1+x_3) ((t_2-t_4) v_F+x_2-x_4) ((t_4-t_3) v_F+x_3+x_4)}
\\&
-\frac{1}{((t_2-t_1) v_F+x_1+x_2) ((t_2-t_3) v_F+x_2-x_3) ((t_1-t_4) v_F-x_1+x_4) ((t_3-t_4) v_F+x_3+x_4)} \bigg)+
\\&
\frac{R^2 T^{*2} }{4 \pi ^4}
 \bigg(-\frac{1}{((t_1-t_2) v_F+x_1+x_2) ((t_1-t_3) v_F+x_1-x_3) ((t_2-t_4) v_F-x_2+x_4) ((t_3-t_4) v_F+x_3+x_4)}
\\&
 -\frac{1}{((t_1-t_2) v_F+x_1+x_2) ((t_2-t_3) v_F-x_2+x_3) ((t_1-t_4) v_F+x_1-x_4) ((t_4-t_3) v_F+x_3+x_4)}
\\&
 -\frac{1}{((t_2-t_1) v_F+x_1+x_2) ((t_1-t_3) v_F-x_1+x_3) ((t_2-t_4) v_F+x_2-x_4) ((t_4-t_3) v_F+x_3+x_4)}
\\&
 -\frac{1}{((t_2-t_1) v_F+x_1+x_2) ((t_2-t_3) v_F+x_2-x_3) ((t_1-t_4) v_F-x_1+x_4) ((t_3-t_4) v_F+x_3+x_4)} \bigg)\\\\
\khatam

\normalsize
\noindent Now it is easy to see that for all the three possible cases above, the fourth moment of density will vanish when either of $|R|$ or $|T|$ vanishes. Hence the Gaussian theory is exact for a homogeneous system and a half line. For all intermediate cases ($0<|R|<1$), the fourth moment (and similarly the higher even moments) are less singular with first order poles as compared to the second moment which contains second order poles. Also note that all the above cases were done for all the four spins to be identical. If any one of them is different, then all the results on the RHSs of the above cases will also vanish.\\

\subsection{Relation between fast and slow parts of DDCF}

\noindent In the RPA sense, the density $ \rho(x,\sigma,t) $ may be ``harmonically analysed" as follows.
\begin{equation}\label{INPUT2}
\rho(x,\sigma,t) = \rho_s(x,\sigma,t) + e^{ 2 i k_F x } \mbox{   }\rho_f(x,\sigma,t) +  e^{ - 2 i k_F x } \mbox{   }\rho^{*}_f(x,\sigma,t)
\end{equation}

\noindent Here $ \rho_s $ and $ \rho_f $ are the slowly varying and the rapidly varying parts respectively.
The most singular parts of the rapidly varying components of the DDCF may be obtained by the following ``non-standard harmonic analysis"
\shuru
\rho_f(x,\sigma,t) \sim e^{ 2 \pi i \int^{x} dy (\rho_s(y,t) + \lambda\mbox{  }\rho_s(-y,t) ) }
\khatam
where $ \lambda = 0 $ or $ \lambda = 1 $ depending upon which correlation function we want to reproduce. The presence of $ \rho_s(-y,t) $ automatically incorporates (at the level of correlation functions), the presence of localised impurities in the system. This is unlike the conventional chiral bosonization method where the impurities lead to non-local hamiltonians whereas the Fermi-Bose correspondence is left untouched. The conventional method, though correct in principle, is unwieldy and unnatural since the impurities which lead to strong qualitative changes in the ground state of the system are introduced as an afterthought whereas the short-range forward scattering which leads only to subtle (but qualitative) changes are treated exactly. NCBT does a good job of including both but it is still not exact as it can only provide the most singular parts of the correlation functions in terms of elementary functions of positions and times.
\\ \mbox{  } \\
Furthermore, in NCBT, the field ``operator" (only a mnemonic for the correlation functions it generates) has a modified form which automatically takes into account the presence of impurities.
\shuru
\\& \psi_{\nu}(x,\sigma,t) \sim e^{ i \Xi_{\nu}(x,\sigma,t)  }  ;
\\& \Xi_{\nu}(x,\sigma,t) \equiv  \theta_{\nu}(x,\sigma,t) + 2 \pi   \lambda \nu \int^{x} dy \mbox{   }\rho_{s}(-y,\sigma,t)
\label{FBCORR}
\khatam
where $ \nu  = R,L $ and $ \lambda = 0,1 $ depending upon which 2-point correlation function we are looking for ( right-right, right-left, left-right, left-left movers and both points on same (opposite) sides of the origin ) .
Also $ \theta_{\nu}(x,\sigma,t) $ is the same as what is found in chiral bosonization. It may be related to currents and densities and through the continuity equation, finally only to the slow parts of the density.
\shuru
 \theta_{\nu}(x,\sigma,t) = \pi \int^{x} dy \left( \nu \rho_s(y,\sigma,t) - \int^{y} dy^{'} \partial_{v_F t} \rho_s(y^{'},\sigma,t) \right)
\khatam
 As usual $  \rho_s(y,\sigma,t) \equiv : \psi^{\dagger}_R(y,\sigma,t)\psi_R(y,\sigma,t) + \psi^{\dagger}_L(y,\sigma,t)\psi_L(y,\sigma,t) :   $
Note that unlike in chiral bosonization, equation (\ref{FBCORR}) is {\it{not}} an operator identity. It is only meant to capture the most singular parts of the two point functions by employing the following device. We assert that the most singular parts of the 2-point functions are given by retaining only the leading terms in the cumulant expansion.\small
\shuru
  <\psi_{\nu}(x,\sigma,t)\psi^{\dagger}_{\nu^{'}}(x^{'},\sigma,t^{'}) >  \mbox{          }  \sim \mbox{    }
   < e^{ i \Xi_{\nu}(x,\sigma,t)  }e^{ - i \Xi_{\nu^{'}}(x^{'},\sigma,t^{'})  }  > \sim e^{ -\frac{1}{2}<\Xi^2_{\nu}(x,\sigma,t)> } e^{ -\frac{1}{2}<\Xi^2_{\nu^{'}}(x^{'},\sigma,t^{'})> }
   e^{  <\Xi_{\nu}(x,\sigma,t)  \Xi_{\nu^{'}}(x^{'},\sigma,t^{'}) > }
   \label{Cumulant}
\khatam \normalsize
As long as the symbol $ < ...  > $ on both sides of the equation (\ref{Cumulant}) is read as ``most singular part of the expectation value", equation (\ref{Cumulant}) is in fact the exact answer for the (most singular part of the) two-point functions of a strongly inhomogeneous Luttinger liquid. The higher order terms in the cumulant expansion are purposely dropped as they can be shown to be less singular (see subsection \ref{Gaussian}).
\\ \mbox{ } \\
\noindent The other important point worth mentioning is that in chiral bosonization, the 2-point Green functions in presence of impurities are discontinuous functions of the short-range forward scattering strength. This means in presence of impurities, the full asymptotic Green functions for weak short-range forward scattering are qualitatively (discontinuously) different from the corresponding quantities for no short-range forward scattering. This is the origin of the ``cutting the chain" and ``healing the chain" metaphors applicable for repulsive and attractive short-range forward scattering respectively.

However NCBT only yields the {\it{most singular}} part of the asymptotic Green functions. This
attribute exhibits a behaviour complementary to the what is seen in the full Green function. The  {\it{ most singular }} part of the 2-point Green functions are discontinuous functions of the {\it{ impurity strength }} in presence of short-range forward scattering between fermions. This means the NCBT Green functions with weak impurities are qualitatively (ie. discontinuously) different from  the corresponding quantity for no impurities.

This is the main reason why the results of chiral bosonization and NCBT cannot be easily compared with one another as they are fully complementary. The only exception is for a fully homogeneous system or its antithesis viz. the half line where the two results coincide.

\subsection{Obtaining the spinless results}

It is easy to transform equation (\ref{full1}) and equation (\ref{full2}) to show the results for the spinless fermions. For doing so, the density density correlation functions of the holons $\langle  \rho_h\rho_h\rangle$ needs to be doubled and that of the spinons $\langle  \rho_n\rho_n\rangle$ are allowed to vanish. The holon velocity in this case will be related to the Fermi velocity as $v_h=\sqrt{v_F^2+v_0 v_F/\pi}$.  Hence the density density correlation function for the spinless case is the following.

\footnotesize
\shuru
\langle T\mbox{   } \rho_s(x_1,t_1)\rho_s(x_2,t_2)\rangle  = &\frac{v_F  }{ 4\pi^2 v_h } \mbox{   } \sum_{  \nu = \pm 1 }\bigg ( -  \frac{1}{ ( x_1-x_2 + \nu v_h(t_1-t_2) )^2 }
	- \frac{ |R|^2 }{    \bigg( 1 -  \frac{(v_h-v_F)}{ v_h } |R|^2   \bigg) }  \frac{\frac{v_F }{v_h}  \mbox{    } \text{sgn}(x_1) \text{sgn}(x_2)\mbox{   }}{  ( | x_1|+|x_2 | + \nu v_h(t_1-t_2) )^2 }
\bigg)\\
\label{spinless}
\khatam

\section{Comparison with perturbative results}

\normalsize
\noindent The density density correlation functions of a strongly inhomogeneous Luttinger liquid given by equation (\ref{full1}) can be verified by a comparison with those obtained using standard fermionic perturbation (the comparison for the limiting cases, viz., $|R|=0$ and $|R|=1$ are given in \hyperref[AppendixB]{Appendix B} and \hyperref[AppendixC]{Appendix C} respectively). The general case viz. $ 0 < |R|  < 1 $ is given in  \hyperref[AppendixD]{Appendix D}. Since the spinon DDCF is the same as that of the non-interacting DDCF (it is only the total density that couples with the interaction), hence it will suffice only to compare the holon DDCF to perturbative results. For this, the holon DDCF is expanded in powers of the interaction parameter $v_0$. Note that the holon velocity $v_h$ is related to the Fermi velocity and the interaction parameter $v_0$ by the relation $v_h=v_F\sqrt{1+2v_0/(\pi v_F)}$.
The holon DDCF is given as follows.
\shuru
\langle T\mbox{   } \rho_h(x_1,t_1)\rho_h(x_2,t_2)\rangle  = &\frac{v_F  }{ 2\pi^2 v_h } \mbox{   } \sum_{  \nu = \pm 1 }\bigg ( -  \frac{1}{ ( x_1-x_2 + \nu v_h(t_1-t_2) )^2 }
	-  \frac{\frac{v_F }{v_h}  \mbox{    } \text{sgn}(x_1x_2)\mbox{   }\frac{ |R|^2 }{    \left( 1 -  \frac{(v_h-v_F)}{ v_h } |R|^2   \right) } }{  ( | x_1|+|x_2 | + \nu v_h(t_1-t_2) )^2 }
\bigg)
\label{fulll1}
\khatam
Expanding in powers of interaction parameter $v_0$ and retaining up to the first order, the following is obtained.
\shuru
\\& \langle T\mbox{   } \rho_h(x_1,t_1)\rho_h(x_2,t_2)\rangle \\& =   \langle T\mbox{   } \rho_h(x_1,t_1)\rho_h(x_2,t_2)\rangle^0  + v_0\mbox{ } \langle T\mbox{   } \rho_h(x_1,t_1)\rho_h(x_2,t_2)\rangle^1 + O[v_0^2]
\khatam
Now
\shuru  \langle T\mbox{   } \rho_h(x_1,t_1)\rho_h(x_2,t_2)\rangle^0 =&  -\frac{1}{2 \pi ^2}\bigg(  \frac{\text{sgn}(x_1) \text{sgn}(x_2) |R|^2}{(\left| x_1\right| +\left| x_2\right| +
v_F (t_1-t_2))^2}+\frac{\text{sgn}(x_1) \text{sgn}(x_2)|R|^2}{(\left| x_1\right| +\left| x_2\right| - v_F(t_1-t_2) )^2} \\&+\frac{1}{(x_1-x_2+v_F(t_1-t_2))^2}+\frac{1}{(x_1-x_2-v_F(t_1-t_2))^2}\bigg)
\khatam
and\small
\shuru
\\&  \langle T\mbox{   } \rho_h(x_1,t_1)\rho_h(x_2,t_2)\rangle^1 = \\& \frac{1}{4 \pi ^3 v_F}
(\frac{|R|^2 \text{sgn}(x_1) \text{sgn}(x_2) (-(|R|^2-1) \left| x_1\right| -(|R|^2-1) \left| x_2\right| +(|R|^2-3) v_F (t_1-t_2))}{(\left| x_1\right| +\left| x_2\right| +v_F (t_2-t_1))^3}\\& +\frac{|R|^2 \text{sgn}(x_1) \text{sgn}(x_2)}{(\left| x_1\right| +\left| x_2\right| +v_F (t_1-t_2))^2}   +\frac{|R|^2 \text{sgn}(x_1) \text{sgn}(x_2)}{(\left| x_1\right| +\left| x_2\right| +v_F (t_2-t_1))^2} \\& -\frac{|R|^2 \text{sgn}(x_1) \text{sgn}(x_2) ((|R|^2-1) \left| x_1\right| +(|R|^2-1) \left| x_2\right| +(|R|^2-3) v_F (t_1-t_2))}{(\left| x_1\right| +\left| x_2\right| +v_F (t_1-t_2))^3} \\& +\frac{2 v_F (t_1-t_2)}{(t_1 v_F-t_2 v_F+x_1-x_2)^3}+\frac{2 v_F (t_1-t_2)}{(t_1 v_F-t_2 v_F-x_1+x_2)^3}+\frac{1}{(t_1 v_F-t_2 v_F+x_1-x_2)^2}+\frac{1}{(t_1 v_F-t_2 v_F-x_1+x_2)^2})\\
\label{firstord}
\khatam\normalsize

\noindent
The first order term viz. equation (\ref{firstord}) has to be compared with that obtained using standard fermionic perturbation theory. Using the perturbative approach, the density density correlation functions in presence of interactions can be written in terms of the non-interacting ones as follows.
\shuru
 \langle T\mbox{   } \rho_h(x_1,t_1)\rho_h(x_2,t_2)\rangle
= \frac{\langle T S \mbox{ } \rho_h(x_1,t_1)\rho_h(x_2,t_2)\rangle_0}{\langle T S \rangle_0}
\khatam
Here $T$ represents the time ordering and the action $S$ can be written as follows.
\shuru
S = e^{-i \int H_{fs}(t)dt} = 1 - i \int H_{fs}(t) dt + ....
\khatam
Hence the zeroth order term is simply $  \langle T\mbox{   } \rho_h(x_1,t_1)\rho_h(x_2,t_2)\rangle_0$  and the first order perturbation term can be written as follows.
\shuruq
  \langle T\mbox{   } \rho_h(x_1,t_1)\rho_h(x_2,t_2)\rangle^1 = -i \int  \langle T\mbox{ } H_I(t)  \rho_h(x_1,t_1)\rho_h(x_2,t_2) \rangle_0 dt
\khatamq
From equation (\ref{FS}) the interacting part of the Hamiltonian can be written as
\shuruq
H_{fs}(t)\mbox{       } = \mbox{         } \frac{1}{2} \int^{ \infty}_{-\infty} dx \int^{\infty}_{-\infty} dx^{'} \mbox{  }v(x-x^{'}) \mbox{   }
 \rho_h(x,t) \rho_h(x^{'},t)
\khatamq
Hence the first order term in the perturbation series  can be written as follows.
\shuru
& \langle T  \rho_h(x_1,t_1)\rho_h(x_2,t_2) \rangle^1 = - \frac{i}{2}
 \int d\tau \mbox{   } \int dy \int dy^{'} \mbox{   }v(y-y^{'})  \langle T \mbox{  } \rho_s(y,\tau_{+})  \rho_s(y^{'},\tau)  \rho_h(x_1,t_1)\rho_h(x_2,t_2) \rangle_0
\label{pert}
\end{aligned}
\end{equation}
Here $v(y-y') = v_0\delta(y-y')$ is the short ranged mutual interaction term. The symbol $\langle..\rangle_0$ on the RHS indicates single particle functions. The next step is crucial. We wish to only include the most singular contributions to the exact first order term in Eq.(\ref{pert}). This involves simply pairing up the densities (as  explained in section \ref{Gaussian}).
Using equation (\ref{pert}), the most singular contribution up to the first order in interaction parameter can be obtained as follows:
\shuru
& \langle T\mbox{   } \rho_h(x_1,t_1)\rho_h(x_2,t_2)\rangle^1  =  -i v_0 \mbox{  }  \int_{C} d\tau\mbox{  } \int_{-\infty}^{\infty} dy \mbox{  }
\langle T\mbox{   }\rho_h(y,\tau) \rho_h(x_1,t_1) \rangle_0\mbox{    } \langle T\mbox{  } \rho_h(y,\tau)\rho_h(x_2,t_2)\rangle_0
\khatam
This has been worked out separately for $ |R| = 0 $, $ |R| = 1 $ and $ 0 < |R| < 1 $ in \hyperref[AppendixB]{Appendix B}, \hyperref[AppendixC]{Appendix C} and \hyperref[AppendixD]{Appendix D} respectively.  In \hyperref[AppendixD]{Appendix D}, the perturbation series to all orders is exhibited in momentum and frequency space and is explicitly evaluated up to second order in $ v_0 $.


\section{Conclusions}

\noindent  In this work, the most singular contributions of the slow part of the density density correlation functions of a Luttinger liquid with short-range forward scattering mutual interactions in presence of static scalar impurities has been rigorously shown to be expressible in terms of elementary functions of positions and times. This formula has simple second order poles (when the 2-point functions are seen as functions of the complex variable $ \tau = t-t^{'}$). Furthermore, it only involves the bare reflection and transmission coefficients of a single fermion in presence of the localised impurities.

 In chiral bosonization, the 2-point Green functions in presence of impurities are discontinuous functions of the short-range forward scattering strength. This means in presence of impurities, the full asymptotic Green functions for weak short-range forward scattering are qualitatively (discontinuously) different from the corresponding quantities for no short-range forward scattering. This is the origin of the ``cutting the chain" and ``healing the chain" metaphors applicable for repulsive and attractive short-range forward scattering respectively.

However NCBT only yields the {\it{most singular}} part of the asymptotic Green functions. This
attribute exhibits a behaviour complementary to the what is seen in the full Green function. The  {\it{ most singular }} part of the 2-point Green functions are discontinuous functions of the {\it{ impurity strength }} in presence of short-range forward scattering between fermions. This means the NCBT Green functions with weak impurities are qualitatively (ie. discontinuously) different from  the corresponding quantity for no impurities.

This is the main reason why the results of chiral bosonization and NCBT cannot be easily compared with one another as they are fully complementary. The only exception is for a fully homogeneous system or its antithesis viz. the half line where the two results coincide.


\section*{APPENDIX  A:   Fourier transform }
\label{AppendixA}
\setcounter{equation}{0}
\renewcommand{\theequation}{A.\arabic{equation}}

\noindent In the main text, we are required to show that equation (\ref{RHORHO0}) may be recovered from equation (\ref{TRNS}) and equation (\ref{DDCFmf}). For this we are required to make sense of integrals such as ($ x \neq x^{'} $ and $ x , x^{'} \neq 0 $),
\shuru
I_0(x-x^{'}) = \sum_{q} e^{ -i q (x-x^{'}) } \mbox{   } \frac{2 q^2 v_F }{w_n^2 + (q v_F)^2}
\khatam
and
\shuru
I_1(x) = \sum_{q} e^{ -i q x } \mbox{   } \frac{2   v_F q }{w_n^2 + (q v_F)^2}
\khatam
We write,
\shuru
I_0(x-x^{'}) = \sum_{q} e^{ -i q (x-x^{'}) } \mbox{   }  \mbox{   }( \frac{ q  }{iw_n  + (q v_F) } + \frac{q  }{-iw_n  + (q v_F) })
\khatam
Set $ X = x-x^{'} $. Then,
\shuru
I_0(X)  = i\frac{d}{dX} \sum_{q} e^{ -i q X } \mbox{   }  \mbox{   }( \frac{ 1  }{iw_n  + (q v_F) } + \frac{1 }{-iw_n  + (q v_F) })  = i\frac{d}{dX}I_1(X)
\khatam
and
\shuru
I_1(x) = \sum_{q} e^{ -i q x } \mbox{   }  \mbox{   }( \frac{ 1  }{iw_n  + (q v_F) } + \frac{1 }{-iw_n  + (q v_F) })
 = -\frac{i L \mbox{      } \text{sgn}(x) \mbox{      } e^{-\frac{\left| w_n\right| \mbox{      }  \left| x\right| }{v_F}}}{v_F}
\khatam
Now,
\shuru
I_1(X > 0)  = -\frac{i L \mbox{      }  e^{-\frac{\left| w_n\right| \mbox{      } X }{v_F}}}{v_F}
\mbox{       }; \mbox{          }\mbox{          }\mbox{          }
I_1(X < 0)  =  \frac{i L \mbox{      }   e^{ \frac{\left| w_n\right| \mbox{      }  X }{v_F}}}{v_F}
\khatam
and
\shuru
I_0(X > 0)    = -   \frac{\left| w_n\right|   }{v_F}   \frac{  L \mbox{      }  e^{-\frac{\left| w_n\right| \mbox{      } X }{v_F}}}{v_F}\mbox{       }; \mbox{          }\mbox{          }\mbox{          }
I_0(X < 0)    = - \frac{\left| w_n\right| }{v_F} \frac{  L \mbox{      }   e^{ \frac{\left| w_n\right| \mbox{      }  X }{v_F}}}{v_F}
\khatam
 or
 \shuru
I_0(X  )    = -   \frac{L   \left| w_n\right|   }{v_F^2}     e^{-\frac{\left| w_n\right| \mbox{      } |X| }{v_F}}
 \khatam
Finally we are called upon to transform the Matsubara frequency to (imaginary) time.
\shuru
J_0(x)   =  \sum_{n} e^{ - w_n \tau } \mbox{     } I_0(x)\mbox{       }; \mbox{          }\mbox{          }\mbox{          }
J_1(x)   =  \sum_{n} e^{ - w_n \tau } \mbox{     } I_1(x)
\khatam
or,
\shuruq
J_0(X)   =   - \sum_{n} e^{ - w_n \tau } \mbox{     } \frac{L   \left| w_n\right|   }{v_F^2}     e^{-\frac{\left| w_n\right| \mbox{      } |X| }{v_F}}  \mbox{       }; \mbox{          }\mbox{          }\mbox{          }
J_1(x)   =  - \sum_{n} e^{ - w_n \tau } \mbox{     } \frac{i L \mbox{      } \text{sgn}(x) \mbox{      } e^{-\frac{\left| w_n\right| \mbox{      }  \left| x\right| }{v_F}}}{v_F}
\khatamq
or,
\shuruq
J_0(X)   =    \frac{ \beta }{2\pi} \int_{-\infty}^{0} dw_n \mbox{     }
e^{ - w_n \tau } \mbox{     } \frac{L     w_n    }{v_F^2}     e^{ \frac{  w_n  \mbox{      } |X| }{v_F}}    -\frac{ \beta }{2\pi} \int_{0}^{\infty} dw_n \mbox{     }
e^{ - w_n \tau } \mbox{     } \frac{L   w_n  }{v_F^2}     e^{-\frac{  w_n \mbox{      } |X| }{v_F}}
\khatamq
\shuruq
J_1(x)   =  - \frac{ \beta }{2\pi} \int_{-\infty}^{0} dw_n \mbox{     }  e^{ - w_n \tau } \mbox{     } \frac{i L \mbox{      } \text{sgn}(x) \mbox{      } e^{ \frac{ w_n \mbox{      }  \left| x\right| }{v_F}}}{v_F} - \frac{ \beta }{2\pi} \int_{0}^{\infty} dw_n \mbox{     }  e^{ - w_n \tau } \mbox{     } \frac{i L \mbox{      } \text{sgn}(x) \mbox{      } e^{-\frac{ w_n  \mbox{      }  \left| x\right| }{v_F}}}{v_F}
\khatamq
or,
\shuruq
J_0(X)   =    \frac{ \beta }{2\pi} \int_{-\infty}^{0} dw_n \mbox{     }
e^{ - w_n \tau } \mbox{     } \frac{L     w_n    }{v_F^2}     e^{ \frac{  w_n  \mbox{      } |X| }{v_F}}    -\frac{ \beta }{2\pi} \int_{0}^{\infty} dw_n \mbox{     }
e^{ - w_n \tau } \mbox{     } \frac{L   w_n  }{v_F^2}     e^{-\frac{  w_n \mbox{      } |X| }{v_F}}
\khatamq
\shuruq
J_1(x)   =  - \frac{ \beta }{2\pi} \int_{-\infty}^{0} dw_n \mbox{     }  e^{ - w_n \tau } \mbox{     } \frac{i L \mbox{      } \text{sgn}(x) \mbox{      } e^{ \frac{ w_n \mbox{      }  \left| x\right| }{v_F}}}{v_F} - \frac{ \beta }{2\pi} \int_{0}^{\infty} dw_n \mbox{     }  e^{ - w_n \tau } \mbox{     } \frac{i L \mbox{      } \text{sgn}(x) \mbox{      } e^{-\frac{ w_n  \mbox{      }  \left| x\right| }{v_F}}}{v_F}
\khatamq
or,
\shuru
J_0(X)   =  -\frac{\beta  L}{2 \pi  (\left| X\right| -\tau  v_F)^2}   -\frac{\beta  L}{2 \pi  (\left| X\right| +\tau  v_F)^2}
\khatam
\shuru
J_1(x)   = \frac{\beta  i L\mbox{     }   \text{sgn}(x)}{2 \pi  (\tau  v_F-\left| x\right| )}-\frac{\beta  i L  \mbox{     } \text{sgn}(x)}{2 \pi  (\left| x\right| +\tau  v_F)}
\khatam
Thus equation (\ref{DENDEN}) is just some combination of these functions  $J_0, J_1 $.

\section*{APPENDIX B:  Perturbative comparison for $|R|=0$ case  }
\label{AppendixB}
\setcounter{equation}{0}
\renewcommand{\theequation}{B.\arabic{equation}}

\shuru
\langle T\mbox{   } \rho_a(x_1,t_1)\rho_a(x_2,t_2)\rangle  = &\frac{v_F  }{ 2\pi^2 v_a } \mbox{   } \sum_{  \nu = \pm 1 }\bigg ( -  \frac{1}{ ( x_1-x_2 + \nu v_a(t_1-t_2) )^2 }
\bigg)
\label{RHORHO}
\khatam
where $ a = n \mbox{  } or \mbox{  } h $,
\shuru
 v_h = \sqrt{v_F^2+\frac{2v_F v_0}{\pi}}
\khatam

$ v_n = v_F $

\shuru
 \langle T\mbox{   } \rho_n(x_1,t_1)\rho_n(x_2,t_2)\rangle  = &\frac{1}{ 2\pi^2 } \mbox{   }
\sum_{  \nu = \pm 1 }\bigg ( -  \frac{1}{ ( x_1-x_2 + \nu v_F(t_1-t_2) )^2 }
\bigg)
\khatam
and $ \langle T\mbox{   } \rho_n(x_1,t_1)\rho_h(x_2,t_2)\rangle  = 0 $. This means $   \langle T\mbox{   } \rho_n(x_1,t_1)\rho_n(x_2,t_2)\rangle^1  = 0 $.

 \vspace{0.1in}{\it{On the one hand}} \vspace{0.1in}

Simply expanding the full final answer  equation (\ref{RHORHO}) to first power in $ v_0 $ we get,\small
\shuru
&  \langle T\mbox{   } \rho_h(x_1,t_1)\rho_h(x_2,t_2)\rangle^1  = \frac{ v_0 }{ 2 \pi^3 v_F }
 \\&
(\frac{2 v_F (t_1-t_2)}{( (t_1-t_2) v_F+x_1-x_2)^3}+\frac{2 v_F (t_1-t_2)}{( (t_1-t_2) v_F-x_1+x_2)^3}
+\frac{1}{( (t_1 -t_2) v_F+x_1-x_2)^2}+\frac{1}{( (t_1-t_2) v_F-x_1+x_2)^2} )
\label{deltaRHORHO}
\khatam\normalsize

 \vspace{0.1in}{\it{On the other hand}} \vspace{0.1in}

Using standard perturbation and retaining the most singular terms we get,
\shuru
&  \langle T\mbox{   } \rho_h(x_1,t_1)\rho_h(x_2,t_2)\rangle^1  =  (-i)v_0 \mbox{  }  \int_{C} d\tau\mbox{  } \int_{-\infty}^{\infty} dy \mbox{  }
\langle T\mbox{   }\rho_h(y,\tau) \rho_h(x_1,t_1) \rangle_0\mbox{    } \langle T\mbox{  } \rho_h(y,\tau)\rho_h(x_2,t_2)\rangle_0
\khatam
But,
\shuru
 \langle T\mbox{   } \rho_h(x_1,t_1)\rho_h(x_2,t_2)\rangle_0  = &\frac{1}{ 2\pi^2 } \mbox{   }
\sum_{  \nu = \pm 1 }\bigg( -  \frac{1}{ ( x_1-x_2 + \nu v_F(t_1-t_2) )^2 }
\bigg)
\khatam

\noindent Hence,
\shuruq
 \langle T\mbox{   } \rho_h(x_1,t_1)\rho_h(x_2,t_2)\rangle^1  =
&\mbox{ }
(-i)v_0 \mbox{  }  \int_{C} d\tau\mbox{  } \int_{-\infty}^{\infty} dy \mbox{  }\frac{1}{4 \pi ^4 (-v_F (\tau -t_1)-x_1+y)^2 (v_F (\tau -t_2)-x_2+y)^2}
\\&
+(-i)v_0 \mbox{  }  \int_{C} d\tau\mbox{  } \int_{-\infty}^{\infty} dy \mbox{  }\frac{1}{4 \pi ^4 (v_F (\tau -t_1)-x_1+y)^2 (-v_F (\tau -t_2)-x_2+y)^2}
\\&
+(-i)v_0 \mbox{  }  \int_{C} d\tau\mbox{  } \int_{-\infty}^{\infty} dy \mbox{  }\frac{1}{4 \pi ^4 (-v_F (\tau -t_1)-x_1+y)^2 (-v_F (\tau -t_2)-x_2+y)^2}
\\&
+(-i)v_0 \mbox{  }  \int_{C} d\tau\mbox{  } \int_{-\infty}^{\infty} dy \mbox{  }\frac{1}{4 \pi ^4 (v_F (\tau -t_1)-x_1+y)^2 (v_F (\tau -t_2)-x_2+y)^2}
\khatamq
Hence,\small
\shuruq
  \langle T&\mbox{   } \rho_h(x_1,t_1)\rho_h(x_2,t_2)\rangle^1  \\
=
&\mbox{ }
(-i)v_0 \mbox{  }  \int_{C} d\tau \mbox{  }\frac{1}{4 \pi ^4 }
\frac{4 \pi i}{ (-v_F (\tau -t_1)-x_1 -    v_F (\tau -t_2)+x_2 )^3 } ( \theta(- (\tau -t_1)) \theta(-(\tau -t_2)) -
 \theta(\tau -t_1) \theta(\tau -t_2)  )
\\&
+(-i)v_0 \mbox{  }  \int_{C} d\tau\mbox{  }\frac{1}{4 \pi ^4 }
\frac{4 \pi i }{ (v_F (\tau -t_1)-x_1  + v_F (\tau -t_2) + x_2 )^3 }
( \theta(\tau -t_1) \theta (\tau -t_2) - \theta( -(\tau -t_1) ) \theta( - (\tau -t_2) )
\\&
+(-i)v_0 \mbox{  }  \int_{C} d\tau \mbox{  }\frac{1}{4 \pi ^4 }
\mbox{    }
 \frac{ 4 \pi i }{ (-v_F (\tau -t_1)-x_1  + v_F (\tau -t_2) + x_2 )^3 } ( \theta(- (\tau -t_1) ) \theta(\tau -t_2)
  - \theta(\tau -t_1) \theta(-(\tau -t_2)) )
\\&
+(-i)v_0 \mbox{  }  \int_{C} d\tau\mbox{  }\frac{1}{4 \pi ^4 }
\mbox{  }
 \frac{ 4 \pi i }{ (v_F (\tau -t_1)-x_1 - v_F (\tau -t_2)+x_2  )^3 }
 (  \theta(\tau -t_1) \theta(-(\tau -t_2)) -  \theta(-(\tau -t_1)) \theta(\tau -t_2) )\\\\
\khatamq\normalsize

\normalsize
\noindent Specifically consider $ t_1 > t_2 $ ($ t_1 $ is on the lower contour and $ t_2 $ is on the upper contour). In this case,
\shuruq
& \theta(- (\tau -t_1)) \theta(-(\tau -t_2)) -
 \theta(\tau -t_1) \theta(\tau -t_2)   =  \theta(-(\tau -t_2)) -
 \theta(\tau -t_1)
\\&
 \theta(\tau -t_1) \theta (\tau -t_2) - \theta( -(\tau -t_1) ) \theta( - (\tau -t_2)  )
  =  \theta(\tau -t_1) - \theta( - (\tau -t_2)  )
\\&
 \theta(- (\tau -t_1) ) \theta(\tau -t_2)
  - \theta(\tau -t_1) \theta(-(\tau -t_2))  = \theta(- (\tau -t_1) ) \theta(\tau -t_2)
\\&
   \theta(\tau -t_1) \theta(-(\tau -t_2)) -  \theta(-(\tau -t_1)) \theta(\tau -t_2)  =
   -  \theta(-(\tau -t_1)) \theta(\tau -t_2)
\khatamq
or,\small
\shuruq
\langle T\mbox{   } \rho_h(x_1,t_1)\rho_h(x_2,t_2)\rangle^1  =
&\mbox{ }
(-i)v_0 \mbox{  }  \int_{C} d\tau \mbox{  }\frac{1}{4 \pi ^4 }
\frac{4 \pi i}{ (-v_F (\tau -t_1)-x_1 -    v_F (\tau -t_2)+x_2 )^3 } (\theta(-(\tau -t_2)) -
 \theta(\tau -t_1)   )
\\&
+(-i)v_0 \mbox{  }  \int_{C} d\tau\mbox{  }\frac{1}{4 \pi ^4 }
\frac{4 \pi i }{ (v_F (\tau -t_1)-x_1  + v_F (\tau -t_2) + x_2 )^3 }
( \theta(\tau -t_1) - \theta( - (\tau -t_2)  ) )
\\&
+(-i)v_0 \mbox{  }  \int_{C} d\tau \mbox{  }\frac{1}{4 \pi ^4 }
\mbox{    }
 \frac{ 4 \pi i }{ (-v_F (\tau -t_1)-x_1  + v_F (\tau -t_2) + x_2 )^3 } ( \theta(- (\tau -t_1) ) \theta(\tau -t_2) )
\\&
+(-i)v_0 \mbox{  }  \int_{C} d\tau\mbox{  }\frac{1}{4 \pi ^4 }
\mbox{  }
 \frac{ 4 \pi i }{ (v_F (\tau -t_1)-x_1 - v_F (\tau -t_2)+x_2  )^3 }
 (  -  \theta(-(\tau -t_1)) \theta(\tau -t_2)  )
\khatamq\normalsize
or,\small
\shuruq
  \langle T\mbox{   } \rho_h(x_1,t_1)\rho_h(x_2,t_2)\rangle^1  =
&\mbox{ }
(-i)v_0 \mbox{  }  \frac{1}{4 \pi ^4 }(\int_{-\infty}^{t_2} d\tau -
\int_{t_1}^{-\infty} d\tau   )
\frac{4 \pi i}{ (-v_F (\tau -t_1)-x_1 -    v_F (\tau -t_2)+x_2 )^3 }
\\&
+(-i)v_0 \mbox{  }  \frac{1}{4 \pi ^4 }(\int_{t_1}^{-\infty} d\tau - \int_{-\infty}^{t_2} d\tau )
\frac{4 \pi i }{ (v_F (\tau -t_1)-x_1  + v_F (\tau -t_2) + x_2 )^3 }
\\&
+(-i)v_0 \mbox{  }  \frac{1}{4 \pi^4 } \int_{t_2}^{t_1} d\tau
\mbox{    }
 \frac{ 4 \pi i }{ (-v_F (\tau -t_1)-x_1  + v_F (\tau -t_2) + x_2 )^3 }
\\&
+(-i)v_0 \mbox{  }  \frac{1}{4 \pi ^4 } (  - \int_{t_2}^{t_1} d\tau  )
\mbox{  }
 \frac{ 4 \pi i }{ (v_F (\tau -t_1)-x_1 - v_F (\tau -t_2)+x_2  )^3 }
\khatamq\normalsize
Or,
\shuruq
  \langle T\mbox{   } \rho_h(x_1,t_1)\rho_h(x_2,t_2)\rangle^1  =
&\mbox{ }
(-i)v_0 \mbox{  }  \frac{1}{4 \pi ^4 }
(\frac{i \pi }{v_F (t_1 v_F-t_2 v_F+x_1-x_2)^2}+\frac{i \pi }{v_F (t_1 v_F-t_2 v_F-x_1+x_2)^2})
\\&
+(-i)v_0 \mbox{  }  \frac{1}{4 \pi ^4 }
(\frac{i \pi }{v_F (t_1 v_F-t_2 v_F+x_1-x_2)^2}+\frac{i \pi }{v_F (t_1 v_F-t_2 v_F-x_1+x_2)^2})
\\&
+(-i)v_0 \mbox{  }  \frac{1}{4 \pi ^4 }
\mbox{    }
 \frac{ 4 \pi i }{ ( x_2-x_1 +  v_F (t_1 -t_2))^3 }
 (t_1-t_2)
\\&
+(-i)v_0 \mbox{  }  \frac{1}{4 \pi ^4 }
\mbox{  }
 \frac{ 4 \pi i }{ ( x_2-x_1 - v_F (t_1 -t_2) )^3 }
 (  -(t_1-t_2)  )
\khatamq
Or,\small
\shuru
& \langle T\mbox{   } \rho_h(x_1,t_1)\rho_h(x_2,t_2)\rangle^1  = \frac{ v_0 }{ 2 \pi^3 v_F }
 \\&
\left(\frac{2 v_F (t_1-t_2)}{( (t_1-t_2) v_F+x_1-x_2)^3}+\frac{2 v_F (t_1-t_2)}{( (t_1-t_2) v_F-x_1+x_2)^3}
+\frac{1}{( (t_1 -t_2) v_F+x_1-x_2)^2}+\frac{1}{( (t_1-t_2) v_F-x_1+x_2)^2} \right)
\label{deltaRHORHO2}
\khatam\normalsize
equation (\ref{deltaRHORHO2}) matches with the earlier result equation (\ref{deltaRHORHO}).

\section*{APPENDIX C:  Perturbative comparison for $|R|=1$ case  }
\label{AppendixC}
\setcounter{equation}{0}
\renewcommand{\theequation}{C.\arabic{equation}}

\begin{equation}
\langle T\mbox{   } \rho_h(x_1,t_1)\rho_h(x_2,t_2)\rangle  = \frac{v_F  }{ 2\pi^2 v_h } \mbox{   } \sum_{  \nu = \pm 1 }\bigg ( -  \frac{1}{ ( x_1-x_2 + \nu v_h(t_1-t_2) )^2 }   -  \frac{\frac{v_F }{v_h}  \mbox{    } \text{sgn}(x_1) \text{sgn}(x_2)\mbox{   }Z_h}{  ( | x_1|+|x_2 | + \nu v_h(t_1-t_2) )^2 }
\bigg)
\end{equation}
\shuru
 Z_h = \frac{ 1 }{    \bigg( 1 - \frac{4 v_0 v_F}{ (v_h+v_F) v_h }
 \mbox{   }
  \frac{ 1 }{    (2\pi) }  \bigg) }\mbox{   } ; \mbox{   }
  v_h = \sqrt{ v_F^2 + \frac{2 v_0 v_F }{\pi} }
\khatam

Expanding to first power of $ v_0 $ we get,

\footnotesize
\shuruq
\langle T&\mbox{   } \rho_h(x_1,t_1)\rho_h(x_2,t_2)\rangle \approx
\\
& -\frac{1}{2 \pi ^2}( \frac{\text{sgn}(x_1) \text{sgn}(x_2)}{(\left| x_1\right| +\left| x_2\right| +
v_F (t_1-t_2))^2}+\frac{\text{sgn}(x_1) \text{sgn}(x_2)}{(\left| x_1\right| +\left| x_2\right| - v_F(t_1-t_2) )^2}
 +\frac{1}{(x_1-x_2+v_F(t_1-t_2))^2}+\frac{1}{(x_1-x_2-v_F(t_1-t_2))^2})
\\&
+ \frac{v_0}{2 \pi^3 v_F} \bigg(\text{sgn}(x_1) \text{sgn}(x_2) \Big(\frac{1}{(\left| x_1\right| +\left| x_2\right|
+ v_F(t_1-t_2))^2}+\frac{1}{(\left| x_1\right| +\left| x_2\right| -v_F(t_1-t_2))^2}\\
&\hspace{4cm}
+ \frac{2v_F(t_1-t_2)}{(\left| x_1\right| +\left| x_2\right| +v_F(t_1-t_2))^3}-\frac{2v_F(t_1-t_2)}{(\left| x_1\right| +\left| x_2\right| -v_F(t_1-t_2))^3}\Big)
\\&
- \frac{2v_F (t_1-t_2)}{(x_1-x_2-v_F(t_1-t_2))^3}+\frac{2v_F (t_1-t_2)}{( x_1-x_2+v_F(t_1-t_2))^3}
+\frac{1}{ (x_1-x_2+v_F(t_1-t_2))^2}+\frac{1}{ (x_1-x_2-v_F(t_1-t_2))^2}\bigg)
\khatamq\normalsize
Or,\footnotesize
\shuru
& \delta\langle T\mbox{   } \rho_h(x_1,t_1)\rho_h(x_2,t_2)\rangle \approx \\& \theta(x_1x_2) \frac{v_0}{2 \pi^3 v_F}
 ( \frac{1}{( x_1 + x_2
+ v_F(t_1-t_2))^2}+\frac{1}{( x_1 + x_2 -v_F(t_1-t_2))^2}
+\frac{1}{ (x_1-x_2+v_F(t_1-t_2))^2}+\frac{1}{ (x_1-x_2-v_F(t_1-t_2))^2})
 \\
+&\theta(x_1x_2)\frac{v_0}{2 \pi^3 v_F}( \frac{2v_F(t_1-t_2)}{( x_1+x_2 +v_F(t_1-t_2))^3}-\frac{2v_F(t_1-t_2)}{( x_1+x_2 -v_F(t_1-t_2))^3}
- \frac{2v_F (t_1-t_2)}{(x_1-x_2-v_F(t_1-t_2))^3}+\frac{2v_F (t_1-t_2)}{( x_1-x_2+v_F(t_1-t_2))^3})
\label{RHOHRHOH1}
\khatam\normalsize
On the other hand, using standard perturbation and retaining the most singular terms we get,
\shuru
&  \langle T\mbox{   } \rho_h(x_1,t_1)\rho_h(x_2,t_2)\rangle^1  =  (-i)v_0 \mbox{  }  \int_{C} d\tau\mbox{  } \int_{-\infty}^{\infty} dy \mbox{  }
\langle T\mbox{   }\rho_h(y,\tau) \rho_h(x_1,t_1) \rangle_0\mbox{    } \langle T\mbox{  } \rho_h(y,\tau)\rho_h(x_2,t_2)\rangle_0
\khatam
This means,\footnotesize

\begin{equation*}
\hspace{-1cm}
\begin{aligned}
 \langle T&\mbox{   } \rho_h(x_1,t_1)\rho_h(x_2,t_2)\rangle^1
=  (-i)v_0   \int_{C} d\tau \int_{-\infty}^{0} dy \mbox{  }
\\
& \frac{1}{2 \pi ^2}\bigg( \frac{-\text{sgn}(x_1)}{(- y +\left| x_1\right| +
v_F (\tau-t_1))^2}+\frac{-\text{sgn}(x_1)}{(- y +\left| x_1\right| - v_F(\tau-t_1) )^2}
 +\frac{1}{(y-x_1+v_F(\tau-t_1))^2}+\frac{1}{(y-x_1-v_F(\tau-t_1))^2}\bigg)
\\\hspace{1cm}&
\mbox{  }
 \frac{1}{2 \pi ^2}\bigg( \frac{- \text{sgn}(x_2)}{(- y  +\left| x_2\right| +
v_F (\tau-t_2))^2}+\frac{- \text{sgn}(x_2)}{( -y  +\left| x_2\right| - v_F(\tau-t_2) )^2}
 +\frac{1}{(y-x_2+v_F(\tau-t_2))^2}+\frac{1}{(y-x_2-v_F(\tau-t_2))^2}\bigg)
\\
&\hspace{3cm}+ (-i)v_0 \mbox{  }  \int_{C} d\tau\mbox{  } \int_{0}^{\infty} dy \mbox{  }\\
& \frac{1}{2 \pi ^2}\bigg( \frac{ \text{sgn}(x_1)}{(y +\left| x_1\right| +
v_F (\tau-t_1))^2}+\frac{ \text{sgn}(x_1)}{( y +\left| x_1\right| - v_F(\tau-t_1) )^2}
 +\frac{1}{(y-x_1+v_F(\tau-t_1))^2}+\frac{1}{(y-x_1-v_F(\tau-t_1))^2}\bigg)
\\\hspace{1cm}&
\mbox{  }
 \frac{1}{2 \pi ^2}\bigg( \frac{\text{sgn}(x_2)}{(y +\left| x_2\right| +
v_F (\tau-t_2))^2}+\frac{\text{sgn}(x_2)}{(y +\left| x_2\right| - v_F(\tau-t_2) )^2}
 +\frac{1}{(y-x_2+v_F(\tau-t_2))^2}+\frac{1}{(y-x_2-v_F(\tau-t_2))^2}\bigg)
\end{aligned}
\end{equation*}
\\ \mbox{       } \\\normalsize
This means,\footnotesize
\begin{equation*}
\hspace{-2cm}
\begin{aligned}
 &\langle T\mbox{   } \rho_h(x_1,t_1)\rho_h(x_2,t_2)\rangle^1
=  (-i)v_0 \mbox{  } \theta(-x_1) \theta(-x_2)\mbox{  } \int_{C} d\tau\mbox{  } \int_{-\infty}^{0} dy \mbox{  }
\\& \frac{1}{2 \pi ^2}\bigg( \frac{1}{(- y - x_1 +
v_F (\tau-t_1))^2}+\frac{1}{(- y - x_1 - v_F(\tau-t_1) )^2}
 +\frac{1}{(y-x_1+v_F(\tau-t_1))^2}+\frac{1}{(y-x_1-v_F(\tau-t_1))^2}\bigg)
\\\hspace{1cm}&
\mbox{  }
 \frac{1}{2 \pi ^2}\bigg( \frac{1}{(- y - x_2 +
v_F (\tau-t_2))^2}+\frac{1}{( -y  - x_2 - v_F(\tau-t_2) )^2}
 +\frac{1}{(y-x_2+v_F(\tau-t_2))^2}+\frac{1}{(y-x_2-v_F(\tau-t_2))^2}\bigg)
\\&\hspace{3cm}
+ (-i)v_0 \mbox{  } \theta(x_1) \theta(x_2)  \mbox{   } \int_{C} d\tau\mbox{  } \int_{0}^{\infty} dy \mbox{  }\\
&
 \frac{1}{2 \pi ^2}\bigg( \frac{1}{(y +x_1 +
v_F (\tau-t_1))^2}+\frac{1}{( y +x_1 - v_F(\tau-t_1) )^2}
 +\frac{1}{(y-x_1+v_F(\tau-t_1))^2}+\frac{1}{(y-x_1-v_F(\tau-t_1))^2}\bigg)
\\\hspace{1cm}&
\mbox{  }
 \frac{1}{2 \pi ^2}\bigg( \frac{1}{(y +x_2 +
v_F (\tau-t_2))^2}+\frac{1}{(y + x_2 - v_F(\tau-t_2) )^2}
 +\frac{1}{(y-x_2+v_F(\tau-t_2))^2}+\frac{1}{(y-x_2-v_F(\tau-t_2))^2}\bigg)
\end{aligned}
\end{equation*}
\\ \mbox{       } \\\normalsize
This means,\footnotesize
\begin{equation*}
\begin{aligned}
  \langle T&\mbox{   } \rho_h(x_1,t_1)\rho_h(x_2,t_2)\rangle^1
=   (-i)v_0 \mbox{  } \theta(x_1 x_2) \mbox{  }\partial_{x_1} \partial_{x_2}\int_{C} d\tau\mbox{  } \int^{0}_{-\infty} dy \mbox{  }
 \frac{1}{4 \pi^4}\\&\times( \frac{1}{(y -x_1 -
v_F (\tau-t_1))}+\frac{1}{(y -x_1 + v_F(\tau-t_1) )}
 -\frac{1}{(y+x_1-v_F(\tau-t_1))}-\frac{1}{(y+x_1+v_F(\tau-t_1))})
\\&
\times( \frac{1}{(y -x_2 -
v_F (\tau-t_2))}+\frac{1}{(y - x_2 + v_F(\tau-t_2) )}
 -\frac{1}{(y+x_2-v_F(\tau-t_2))}-\frac{1}{(y+x_2+v_F(\tau-t_2))})
\end{aligned}
\end{equation*}
\normalsize
This means,\small
\shuruq
  \langle T\mbox{   }& \rho_h(x_1,t_1)\rho_h(x_2,t_2)\rangle^1\\
=& \sum_{ a_1,\nu_1,a_2, \nu_2 = \pm 1}  (-i)v_0 \mbox{  } \theta(x_1 x_2) \mbox{  }\partial_{x_1} \partial_{x_2}\int_{C} d\tau\mbox{  } \int^{0}_{-\infty} dy \mbox{  }
 \frac{1}{4 \pi^4}\mbox{             } \frac{a_1}{(y -a_1 x_1 -\nu_1
v_F (\tau-t_1))} \frac{a_2}{(y -a_2 x_2 -\nu_2
v_F (\tau-t_2))}
\khatamq
\\ \mbox{       } \\
\normalsize This also means,\small
\shuruq
  \langle T&\mbox{   } \rho_h(x_1,t_1)\rho_h(x_2,t_2)\rangle^1
\\
=& \sum_{ a_1,\nu_1,a_2, \nu_2 = \pm 1}  (-i)v_0 \mbox{  } \theta(x_1 x_2) \mbox{  }\partial_{x_1} \partial_{x_2}\int_{C} d\tau\mbox{  } \int_{0}^{\infty} dy \mbox{  }
 \frac{1}{4 \pi^4}\mbox{             } \frac{a_1}{(y -a_1 x_1 -\nu_1
v_F (\tau-t_1))} \frac{a_2}{(y -a_2 x_2 -\nu_2
v_F (\tau-t_2))}
\khatamq
\\ \mbox{       } \\
\normalsize This means,\small
\shuruq
 \langle T&\mbox{   } \rho_h(x_1,t_1)\rho_h(x_2,t_2)\rangle^1
\\&
= \sum_{ a_1,\nu_1,a_2, \nu_2 = \pm 1}  (-i)\frac{v_0}{2} \mbox{  } \theta(x_1 x_2) \mbox{  }\int_{C} d\tau\mbox{  } \int^{\infty}_{-\infty} dy \mbox{  }
 \frac{1}{4 \pi^4}\mbox{             } \frac{1}{(y -a_1 x_1 -\nu_1
v_F (\tau-t_1))^2 (y -a_2 x_2 -\nu_2
v_F (\tau-t_2))^2}
\khatamq
\\ \mbox{       } \\
\normalsize This means, \small
\shuru
 \langle T&\mbox{   } \rho_h(x_1,t_1)\rho_h(x_2,t_2)\rangle^1
\\&
= \sum_{ a_1,\nu_1,a_2, \nu_2 = \pm 1}  (-i)\frac{v_0}{2} \mbox{  } \theta(x_1 x_2) \mbox{  }\int_{C} d\tau\mbox{  } \int^{\infty}_{-\infty} dy \mbox{  }
 \frac{1}{4 \pi^4}\mbox{             } \frac{1}{(y -a_1 x_1 -\nu_1
v_F (\tau-t_1))^2 (y -a_2 x_2 -\nu_2
v_F (\tau-t_2))^2}
\khatam
\\ \mbox{       } \\
\shuru
  \langle T\mbox{   }& \rho_h(x_1,t_1)\rho_h(x_2,t_2)\rangle^1
\\&
= \sum_{ a_1,\nu_1,a_2, \nu_2 = \pm 1}  (-i)\frac{v_0}{2} \mbox{  } \theta(x_1 x_2) \mbox{  }\int_{C} d\tau\mbox{  }
  \mbox{  }
 \frac{1}{4 \pi^4}\mbox{             }
\frac{ 4 \pi i }{ (-a_1 x_1 -\nu_1v_F (\tau-t_1) -( -a_2 x_2 -\nu_2v_F (\tau-t_2)))^3 }
\\&\hspace{2cm}
\times (\theta(-Im[ -\nu_1(\tau-t_1)])\theta(Im[-\nu_2 (\tau-t_2)])-\theta(Im[ -\nu_1(\tau-t_1)])\theta(-Im[-\nu_2 (\tau-t_2)]) )
\khatam
\\ \mbox{       } \\
\shuruq
  \langle T\mbox{   } \rho_h(x_1,t_1)\rho_h(x_2,t_2)\rangle^1&\\
= \sum_{ a_1,\nu_1,a_2, \nu_2 = \pm 1}&  (-i)\frac{v_0}{2} \mbox{  } \theta(x_1 x_2) \mbox{  }\int_{C} d\tau\mbox{  }
  \mbox{  }
 \frac{1}{4 \pi^4}\mbox{             }
\frac{ 4 \pi i }{ (-a_1 x_1 -\nu_1v_F (\tau-t_1) -( -a_2 x_2 -\nu_2v_F (\tau-t_2)))^3 }\mbox{  } \mbox{  }\\
&(\theta_C( -\nu_1(\tau-t_1))\theta_C(\nu_2 (\tau-t_2))-\theta_C( \nu_1(\tau-t_1))\theta_C(-\nu_2 (\tau-t_2)) )
\khatamq
\\ \mbox{       } \\\small
\shuruq
  \langle T&\mbox{   } \rho_h(x_1,t_1)\rho_h(x_2,t_2)\rangle^1
\\&
= \sum_{ a_1,a_2,\nu  = \pm 1}  (-i)\frac{v_0}{2} \mbox{  } \theta(x_1 x_2) \mbox{  }\int_{C} d\tau\mbox{  }
  \mbox{  }
 \frac{1}{4 \pi^4}\mbox{             }
\frac{ 4 \pi i }{ (-a_1 x_1+a_2 x_2  + \nu v_F (t_1-t_2)  )^3 }\mbox{  } \mbox{  }\\
&\hspace{2cm}(\theta_C( -\nu (\tau-t_1))\theta_C(\nu (\tau-t_2))-\theta_C( \nu (\tau-t_1))\theta_C(-\nu (\tau-t_2)) )
\\&
+ \sum_{ a_1,a_2, \nu = \pm 1}  (-i)\frac{v_0}{2} \mbox{  } \theta(x_1 x_2) \mbox{  }\int_{C} d\tau\mbox{  }
  \mbox{  }
 \frac{1}{4 \pi^4}\mbox{             }
\frac{ 4 \pi i }{ (-a_1 x_1 +a_2 x_2 + \nu v_F (t_1+t_2) -2\nu v_F \tau )^3 }\mbox{  } \mbox{  }\\
&\hspace{2cm}(\theta_C( -\nu (\tau-t_1))\theta_C(- \nu  (\tau-t_2))-\theta_C( \nu (\tau-t_1))\theta_C( \nu (\tau-t_2)) )
\khatamq
\\ \mbox{       } \\\small
\shuruq
  \langle& T\mbox{   } \rho_h(x_1,t_1)\rho_h(x_2,t_2)\rangle^1= \sum_{ a_1,a_2 = \pm 1}  (-i)\frac{v_0}{2} \mbox{  } \theta(x_1 x_2)
\\
& \bigg(\mbox{  }\int_{C} d\tau\mbox{  }
  \mbox{  }
 \frac{1}{4 \pi^4}\mbox{             }
\frac{ 4 \pi i }{ (-a_1 x_1+a_2 x_2  +  v_F (t_1-t_2)  )^3 }\mbox{  } \mbox{  }
(\theta_C( - (\tau-t_1))\theta_C(\tau-t_2)-\theta_C(\tau-t_1)\theta_C(-(\tau-t_2)) )
\\&
+ \mbox{  }\int_{C} d\tau\mbox{  }
  \mbox{  }
 \frac{1}{4 \pi^4}\mbox{             }
\frac{ 4 \pi i }{ (-a_1 x_1+a_2 x_2  - v_F (t_1-t_2)  )^3 }\mbox{  } \mbox{  }
(\theta_C(\tau-t_1)\theta_C(- (\tau-t_2))-\theta_C( - (\tau-t_1))\theta_C(\tau-t_2) )
\\&
+ \mbox{  }\int_{C} d\tau\mbox{  }
  \mbox{  }
 \frac{1}{4 \pi^4}\mbox{             }
\frac{ 4 \pi i }{ (-a_1 x_1 +a_2 x_2 - v_F (t_1+t_2) +2 v_F \tau )^3 }\mbox{  } \mbox{  }
(\theta_C(\tau-t_1)\theta_C(\tau-t_2)-\theta_C( -(\tau-t_1))\theta_C( - (\tau-t_2)) )
\\&
+ \mbox{  }\int_{C} d\tau\mbox{  }
  \mbox{  }
 \frac{1}{4 \pi^4}\mbox{             }
\frac{ 4 \pi i }{ (-a_1 x_1 +a_2 x_2 +  v_F (t_1+t_2) -2 v_F \tau )^3 }\mbox{  } \mbox{  }
(\theta_C( - (\tau-t_1))\theta_C(-   (\tau-t_2))-\theta_C(\tau-t_1)\theta_C(\tau-t_2) \bigg)
\khatamq\normalsize
\\ \mbox{       } \\
If $ t_1 > t_2 $,\small
\shuru
 \langle T\mbox{   } \rho_h(x_1&,t_1)\rho_h(x_2,t_2)\rangle^1\\
=& \sum_{ a_1,a_2 = \pm 1}  (-i)\frac{v_0}{2} \mbox{  } \theta(x_1 x_2) \mbox{  }
  \mbox{  }
 \frac{1}{4 \pi^4}\mbox{             }
\frac{ 4 \pi i }{ (-a_1 x_1+a_2 x_2  +  v_F (t_1-t_2)  )^3 }\mbox{  } \mbox{  }
(\int_{t_2}^{t_1} d\tau )
\\&
+ \sum_{ a_1,a_2 = \pm 1}  (-i)\frac{v_0}{2} \mbox{  } \theta(x_1 x_2) \mbox{  }
  \mbox{  }
 \frac{1}{4 \pi^4}\mbox{             }
\frac{ 4 \pi i }{ (-a_1 x_1+a_2 x_2  - v_F (t_1-t_2)  )^3 }\mbox{  } \mbox{  }
(-\int_{t_2}^{t_1} d\tau )
\\&
+ \sum_{ a_1,a_2  = \pm 1 }  (-i)\frac{v_0}{2} \mbox{  } \theta(x_1 x_2) \mbox{  }
  \mbox{  }
 \frac{1}{4 \pi^4}\mbox{             }(-\int^{t_1}_{-\infty} d\tau -\int_{-\infty}^{t_2} d\tau  )
\frac{ 4 \pi i }{ (-a_1 x_1 +a_2 x_2 - v_F (t_1+t_2) +2 v_F \tau )^3 }\mbox{  } \mbox{  }
\\&
+ \sum_{ a_1,a_2 = \pm 1}  (-i)\frac{v_0}{2} \mbox{  } \theta(x_1 x_2) \mbox{  }
  \mbox{  }
 \frac{1}{4 \pi^4}\mbox{             }
(\int_{-\infty}^{t_2} d\tau+\int^{t_1}_{-\infty} d\tau )\frac{ 4 \pi i }{ (-a_1 x_1 +a_2 x_2 +  v_F (t_1+t_2) -2 v_F \tau )^3 }\mbox{  }
\khatam
\shuru
  \langle T\mbox{   } \rho_h(x_1&,t_1)\rho_h(x_2,t_2)\rangle^1\\
= &\sum_{ a_1,a_2 = \pm 1}  (-i)\frac{v_0}{2} \mbox{  } \theta(x_1 x_2) \mbox{  }
  \mbox{  }
 \frac{1}{4 \pi^4}\mbox{             }
\frac{ 4 \pi i }{ (-a_1 x_1+a_2 x_2  +  v_F (t_1-t_2)  )^3 }\mbox{  } \mbox{  }
(t_1-t_2)
\\&
+ \sum_{ a_1,a_2 = \pm 1}  (-i)\frac{v_0}{2} \mbox{  } \theta(x_1 x_2) \mbox{  }
  \mbox{  }
 \frac{1}{4 \pi^4}\mbox{             }
\frac{ 4 \pi i }{ (-a_1 x_1+a_2 x_2  - v_F (t_1-t_2)  )^3 }\mbox{  } \mbox{  }
(t_2-t_1)
\\&
+ \sum_{ a_1,a_2 = \pm 1}  (-i)\frac{v_0}{2} \mbox{  } \theta(x_1 x_2) \mbox{  }
  \mbox{  }
 \frac{1}{4 \pi^4}\mbox{             }(-\int^{t_1}_{-\infty} d\tau -\int_{-\infty}^{t_2} d\tau  )
\frac{ 4 \pi i }{ (-a_1 x_1 +a_2 x_2 - v_F (t_1+t_2) +2 v_F \tau )^3 }\mbox{  } \mbox{  }
\\&
+ \sum_{ a_1,a_2 = \pm 1}  (-i)\frac{v_0}{2} \mbox{  } \theta(x_1 x_2) \mbox{  }
  \mbox{  }
 \frac{1}{4 \pi^4}\mbox{             }
(\int_{-\infty}^{t_2} d\tau+\int^{t_1}_{-\infty} d\tau )\frac{ 4 \pi i }{ (-a_1 x_1 +a_2 x_2 +  v_F (t_1+t_2) -2 v_F \tau )^3 }\mbox{  }
\khatam\footnotesize
\shuru
\\  \langle T\mbox{   }& \rho_h(x_1,t_1)\rho_h(x_2,t_2)\rangle^1
 = \sum_{ a_1,a_2 = \pm 1}  (-i)v_0 \mbox{  } \theta(x_1 x_2) \mbox{  }
  \mbox{  }
 \frac{1}{4 \pi^4}\mbox{             }
\frac{ 4 \pi i }{ (-a_1 x_1+a_2 x_2  +  v_F (t_1-t_2)  )^3 }\mbox{  } \mbox{  }
(t_1-t_2)
\\&
+ \sum_{ a_1,a_2 = \pm 1}  (-i)v_0 \mbox{  } \theta(x_1 x_2) \mbox{  }
  \mbox{  }
 \frac{1}{4 \pi^4}\mbox{             }(\frac{i \pi }{v_F (a_1x_1-a_2x_2+t_1 v_F-t_2 v_F)^2}+\frac{i \pi }{v_F (-a_1x_1+a_2x_2+t_1 v_F-t_2 v_F)^2})
\khatam
\\ \mbox{       } \\
\shuru
  \langle T\mbox{   } \rho_h(x_1,t_1)\rho_h(x_2,t_2)\rangle^1
= \sum_{ a_1,a_2 = \pm 1} \frac{ v_0 }{2\pi^3 v_F} \mbox{  } \theta(x_1 x_2) \mbox{  }
  \mbox{  }
\left( \frac{ 2 v_F (t_1-t_2) }{ (-a_1 x_1+a_2 x_2  +  v_F (t_1-t_2)  )^3 }\mbox{  } \mbox{  }
+ \frac{1}{(a_1x_1-a_2x_2+  v_F(t_1-t_2))^2} \right)
 \mbox{      } \mbox{      } \mbox{           }  \label{RHOHRHOH2}
\khatam
Hence equation (\ref{RHOHRHOH1}) matches with equation (\ref{RHOHRHOH2})

\section*{APPENDIX D:  Perturbative comparison for $0 < |R| <  1$ case  }
\label{AppendixD}
\setcounter{equation}{0}
\renewcommand{\theequation}{D.\arabic{equation}}
\normalsize
\noindent The conventional method of performing perturbation expansion is the S-matrix method viz. $ \rho_s(x_1,t_1) = \rho_s(x_1,\uparrow, t_1) + \rho_s(x_1,\downarrow,t_1) $,
\shuru
\\ <T \mbox{    } \rho_s(x_1,t_1) \rho_s (x_2,t_2)  >
\mbox{    } &=\mbox{    } \frac{<T \mbox{    }S \mbox{  } \rho_s(x_1,t_1) \rho_s(x_2,t_2)  >_0  }{ <T \mbox{    }S   >_0} \\& =\mbox{    } <T \mbox{    } \rho_s(x_1,t_1) \rho_s (x_2,t_2)  >_0 - i \int_Cdt \mbox{   }<T \mbox{    } H_{fs}(t) \mbox{  } \rho_s(x_1,t_1) \rho_s(x_2,t_2)  >_{0,c}
 \\& + \mbox{ }\frac{ (-i)^2 }{2} \int_Cdt \mbox{   }\int_Cdt^{'} \mbox{   }<T \mbox{    } H_{fs}(t) H_{fs}(t^{'})\mbox{  } \rho_s(x_1,t_1) \rho_s (x_2,t_2)  >_{0,c}
\khatam
where $ S = e^{ -i \int_Cdt \mbox{   }  H_{fs}(t) } $ where $ H_{fs} $ is given by equation (\ref{FS}). Retaining only the most singular terms, \footnotesize
\shuru
\\& <T \mbox{    } \rho_s(x_1, t_1) \rho_s (x_2, t_2)  >
  = <T \mbox{    } \rho_s(x_1, t_1) \rho_s (x_2, t_2)  >_0 - iv_0 \int_Cdt \mbox{   }
\int dy  <T \mbox{    }    \rho_s(x_1, t_1)\rho_s(y, t)
>_{0} <T \mbox{    }
\rho_s(y, t)  \rho_s(x_2, t_2)  >_{0}
 \\& +  v_0^2 (-i)^2   \int_Cdt \mbox{   }\int_Cdt^{'} \mbox{   }
 \int dy \mbox{ }
\int dz \mbox{ } <T \mbox{    }  \rho_s(x_1, t_1)  \rho_s(z, t )
>_0\mbox{ } <T \mbox{    }\rho_s(z, t)  \rho_s(y, t^{'})
>_0 \mbox{ }
 <T \mbox{    } \rho_s(y, t^{'}) \rho_s (x_2, t_2)  >_{0} +  ....
\khatam\normalsize
Since we are going beyond leading order, it is more convenient to work in momentum and frequency space.
\shuru
<T \mbox{   }\rho_s(x_1,t_1 )\rho_s(x_{2},t_{2} )>  =
\frac{1}{L^2} \sum_{q,q^{'},n }e^{ - i q x_1 } e^{ - i q^{'} x_2 } e^{ - w_n (t_1-t_2)}
<\rho_{q,n;.}\rho_{q^{'},-n;. }>
 \khatam
Thus the most singular parts are captured by the following perturbation series,
\shuru
\\  <\rho_{q,n;.}\rho_{q^{'},-n;. }>
   &=\mbox{    }
<\rho_{q,n;.}\rho_{q^{'},-n;. }>_0
   -   \frac{  v_0\beta }{L} \sum_{ Q^{'}  }
<\rho_{q,n;.}\rho_{Q^{'},-n;. }>_0
<\rho_{-Q^{'},n;.}\rho_{q^{'},-n;. }>_0
 \\& +
 \frac{ v_0^2 \beta^2}{L^2}
  \sum_{ Q^{'},Q^{''}  }
<\rho_{q,n;.}\rho_{Q^{'},-n; .}>_0
<\rho_{-Q^{'},n; .}\rho_{Q^{''},-n;. }>_0
<\rho_{-Q^{''},n;.}\rho_{q^{'},-n;. }>_0
 +  ....
 \label{series}
\khatam
{\it{ NCBT says that when the above series is summed to all orders we get}},
\shuru
 <\rho_{q,n;.}\rho_{q^{'},-n;.}> =  \delta_{q+q^{'},0}
   \mbox{  }\frac{L}{\beta  }  \frac{  2v_F   q^2  }{   \pi ( w^2_n  + (q  v_h)^2 ) }
+   \frac{ |w_n|   }{\pi \beta} \frac{ |R|^2 }{     \left( 1 -   \frac{(v_h-v_F)}{  v_h }
 \mbox{   }
 |R|^2 \right) } \mbox{  } \frac{   (2 v_F q)(2 v_F q^{'}) }{
  \left(  w^2_n  + (q^{'}  v_h)^2   \right)   \left( w^2_n  + (q  v_h)^2   \right) }
  \label{NCBTRHORHO}
 \\
\khatam
where $ v_h = \sqrt{ v_F^2 + \frac{2 v_0 v_F}{\pi}} $.
However when $ v_h = v_F $ (no short-range forward scattering between fermions),
\shuru
 <\rho_{q,n;.}\rho_{q^{'},-n;.}>_0 \mbox{  }
=  \mbox{  }
\delta_{q+q^{'},0}
   \mbox{  }\frac{L}{\beta  }  \frac{  2v_F   q^2  }{   \pi ( w^2_n  + (q  v_F)^2 ) }
+   \frac{ |w_n|   }{\pi \beta}  |R|^2   \mbox{  } \frac{   (2 v_F q)(2 v_F q^{'}) }{
  \left(  w^2_n  + (q^{'}  v_F)^2   \right)   \left( w^2_n  + (q  v_F)^2   \right) }
 \\
\khatam
Here we want to verify that this is consistent upto second order. The resummation to all orders has already been done using the generating function method in the main text. The series equation (\ref{series}) may be evaluated as follows:
\\ \mbox{   } \\
\noindent {\it{The term proportional to $ v_0 $ in $ <\rho_{q,n;.}\rho_{q^{'},-n;.}>-<\rho_{q,n;.}\rho_{q^{'},-n;.}>_0 $ in the series equation (\ref{series}) is}}:
\shuru
\\&-\frac{4 L q^2 q^{'2} v_0 v_F^2 \delta _{0,q+q^{'}}}{\pi ^2 \beta  \left(q^2 v_F^2+w_n^2\right) \left(q^{'2} v_F^2+w_n^2\right)}+\frac{L \int_{-\infty }^{\infty } \frac{16 q q_1^2 q^{'} |R|^4 v_0 v_F^4 \left| w_n\right| ^2}{\pi ^2 \beta  L \left(q^2 v_F^2+w_n^2\right) \left(q_1^2 v_F^2+w_n^2\right)^2 \left(q^{'2} v_F^2+w_n^2\right)} \, dq_1}{2 \pi }\\& -\frac{8 q q^{'3} |R|^2 v_0 v_F^3 \left| w_n\right| }{\pi ^2 \beta  \left(q^2 v_F^2+w_n^2\right) \left(q^{'2} v_F^2+w_n^2\right)^2}-\frac{8 q^3 q^{'} |R|^2 v_0 v_F^3 \left| w_n\right| }{\pi ^2 \beta  \left(q^2 v_F^2+w_n^2\right)^2 \left(q^{'2} v_F^2+w_n^2\right)}
\khatam
This simplifies to,
\shuru
T_1 = -\frac{4 q q^{'} v_0 v_F \bigg(
\substack{L q q^{'} v_F \left(q^2 v_F^2+w_n^2\right) \left(q^{'2} v_F^2+w_n^2\right) \delta _{0,q+q^{'}}-
|R|^2 \left| w_n\right|  \left(q^2 q^{'2} (|R|^2-4) v_F^4+(|R|^2-2) v_F^2 w_n^2 \left(q^2+q^{'2}\right)+|R|^2 w_n^4\right)}\bigg)}
{\pi ^2 \beta  \left(q^2 v_F^2+w_n^2\right)^2 \left(q^{'2} v_F^2+w_n^2\right)^2}
\khatam
\\ \mbox{   } \\
\noindent {\it{The term proportional to $ v_0^2 $ in $ <\rho_{q,n;.}\rho_{q^{'},-n;.}>-<\rho_{q,n;.}\rho_{q^{'},-n;.}>_0 $ in the series equation (\ref{series}) is}}:\small
\shuru
\\&\frac{8 L q^2 q^4 v_0^2 v_F^3 \delta _{0,q+q^{'}}}{\pi ^3 \beta  \left(q^2 v_F^2+w_n^2\right) \left(q^2 v_F^2+w_n^2\right) \left(q^2 v_F^2+w_n^2\right)}\\
&+ \left( \frac{L}{2\pi} \right)^2 \mbox{ }  \int _{-\infty }^{\infty }\int _{-\infty }^{\infty }\frac{64 q q^{'} Q^{'2} Q^{''2} |R|^6 v_0^2 v_F^6 \left| w_n\right| ^3}{\pi ^3 \beta  L^2 \left(q^2 v_F^2+w_n^2\right) \left(q^{'2} v_F^2+w_n^2\right) \left(Q^{'2} v_F^2+w_n^2\right)^2 \left(Q^{''2} v_F^2+w_n^2\right)^2}dQ^{''}dQ^{'} \\& -\frac{L }{2 \pi }\int_{-\infty }^{\infty } \frac{32 q q^{'} Q^{'3}  Q^{'} |R|^4 v_0^2 v_F^5 \left| w_n\right| ^2}{\pi ^3 \beta  L \left(q^2 v_F^2+w_n^2\right) \left(q^{'2} v_F^2+w_n^2\right) \left(Q^{'2} v_F^2+w_n^2\right)^2 \left(Q^{'2} v_F^2+w_n^2\right)} \, dQ^{'}\\& +\frac{L}{2 \pi } \int_{-\infty }^{\infty } \frac{32 q^2 (-q) q^{'} Q^{''2} |R|^4 v_0^2 v_F^5 \left| w_n\right| ^2}{\pi ^3 \beta  L \left(q^2 v_F^2+w_n^2\right) \left(q^2 v_F^2+w_n^2\right) \left(q^{'2} v_F^2+w_n^2\right) \left(Q^{''2} v_F^2+w_n^2\right)^2} \, dQ^{''}\\& -\frac{L}{2 \pi } \int_{-\infty }^{\infty } \frac{32 q q^{'3} Q^{'2} |R|^4 v_0^2 v_F^5 \left| w_n\right| ^2}{\pi ^3 \beta  L \left(q^2 v_F^2+w_n^2\right) \left(q^{'2} v_F^2+w_n^2\right)^2 \left(Q^{'2} v_F^2+w_n^2\right)^2} \, dQ^{'}\\
&-\frac{16 q^2 (-q)^3 q^{'} |R|^2 v_0^2 v_F^4 \left| w_n\right| }{\pi ^3 \beta  \left(q^2 v_F^2+w_n^2\right) \left(q^2 v_F^2+w_n^2\right) \left(q^2 v_F^2+w_n^2\right) \left(q^{'2} v_F^2+w_n^2\right)}\\
& +\frac{16 q q^{'3} q^{'2} |R|^2 v_0^2 v_F^4 \left| w_n\right| }{\pi ^3 \beta  \left(q^2 v_F^2+w_n^2\right) \left(q^{'2} v_F^2+w_n^2\right)^2 \left(q^{'2} v_F^2+w_n^2\right)}-\frac{16 q^2 (-q) q^{'3} |R|^2 v_0^2 v_F^4 \left| w_n\right| }{\pi ^3 \beta  \left(q^2 v_F^2+w_n^2\right) \left(q^2 v_F^2+w_n^2\right) \left(q^{'2} v_F^2+w_n^2\right)^2}
\khatam\normalsize
This simplifies to,\small
\shuru
\\& T_2 = \frac{2 q q^{'} v_0^2 }{\pi ^3 \beta  (q^2 v_F^2+w_n^2)^3 (q^{'2} v_F^2+w_n^2)^3}  \bigg(4 L q^2 q^{'2} v_F^3 (q^2 v_F^2+w_n^2) (q^{'2} v_F^2+w_n^2)  (q q^{'} v_F^2-w_n^2)\delta _{0,q+q^{'}}\\
&\hspace{ 1cm} +|R|^2 \left| w_n\right|  (q^4 q^{'4} (|R|^2 (2 |R|^2-11)+24) v_F^8+2 q^2 q^{'2} (|R|^2 (2 |R|^2-9)+12) v_F^6 w_n^2 (q^2+q^{'2})\\
&\hspace{ 1cm}  +2 |R|^2 (2 |R|^2-5) v_F^2 w_n^6 (q^2+q^{'2})+v_F^4 w_n^4 ((|R|^2 (2 |R|^2-7)+8) (q^4+q^{'4})\\
&\hspace{ 1cm}+4 q^2 q^{'2} (|R|^2 (2 |R|^2-7)+2))+|R|^2 (2 |R|^2-3) w_n^8)\bigg)
\khatam
\normalsize
It is easy to verify that both $T_1 $ and $T_2 $ may also be obtained by simply expanding equation (\ref{NCBTRHORHO}) in powers of $ v_0 $ and retaining upto order $ v_0^2 $.

\section*{APPENDIX E:  Derivation of formulas for the parameters  of the generalized Hamiltonian ($ V_R,V_L, V_1, V_1^{*} $) in terms of $ T,R $ }
\label{AppendixE}
\setcounter{equation}{0}
\renewcommand{\theequation}{E.\arabic{equation}}

\shuru
H_0 =& -i {\text{ $v_F$ }} \int dx \mbox{ } (\psi^{\dagger}_R(x)\partial_x \psi_R(x) - \psi^{\dagger}_L(x)\partial_x \psi_L(x) )
 + V_R \psi^{\dagger}_R(0)\psi_R(0) + V_L \psi^{\dagger}_L(0)\psi_L(0)\\
&  + V_1 \mbox{  } \psi^{\dagger}_R(0)\psi_L(0)   + V^{*}_1 \mbox{  } \psi^{\dagger}_L(0)\psi_R(0)
\khatam
Note that we may write $ V_1 = |V_1| \mbox{ }e^{ i \delta } $. This phase may be absorbed by a redefinition of the fields $ \psi_R(x) \rightarrow e^{ i \delta } \psi_R(x) $ for example.
This means $ V_1 $ can be chosen to be real without loss of generality.
\shuru
<T \mbox{ } \psi_{\nu}(x,t)  \psi^{\dagger}_{\nu^{'}}(x^{'},t^{'}) > \mbox{  } =  \mbox{  }\frac{ \Gamma^{\nu,\nu^{'}}(x,x^{'}) }{ \nu x - \nu^{'} x^{'} - {\text{ $v_F$ }} (t-t^{'}) }
\khatam
and
\shuru
\Gamma^{\nu,\nu^{'}}(x,x^{'}) = \sum_{ \gamma,\gamma^{'} = \pm 1} \theta(\gamma x) \theta(\gamma^{'} x^{'}) \mbox{    }
g_{\gamma,\gamma^{'}}(\nu,\nu^{'})
\khatam

\shuru
i\partial_t <  \psi_{\nu}(x,t)  \psi^{\dagger}_{\nu^{'}}(x^{'},t^{'}) > \mbox{  } = i{\text{ $v_F$ }}   \mbox{  }\frac{ \Gamma^{\nu,\nu^{'}}(x,x^{'}) }{( \nu x - \nu^{'} x^{'} - {\text{ $v_F$ }} (t-t^{'}) )^2 }
\khatam

\shuru
i\partial_t <  \psi_{\nu}(x,t)  \psi^{\dagger}_{\nu^{'}}(x^{'},t^{'}) > \mbox{  } =  \mbox{  }
<  [\psi_{\nu}(x,t) , H_0] \psi^{\dagger}_{\nu^{'}}(x^{'},t^{'}) >
\khatam\normalsize
or\small
\shuruq
 i\partial_t <  \psi_{\nu}(x,t)  \psi^{\dagger}_{\nu^{'}}(x^{'},t^{'}) > \mbox{  } &=   \mbox{ }  -i {\text{ $v_F$ }}  \nu\mbox{ }
<    \partial_x \psi_{\nu}(x,t) \psi^{\dagger}_{\nu^{'}}(x^{'},t^{'}) >
\\&
+ <  [\psi_{\nu}(x,t) ,  V_R \psi^{\dagger}_R(0)\psi_R(0)  ] \psi^{\dagger}_{\nu^{'}}(x^{'},t^{'}) >
  + <  [\psi_{\nu}(x,t) ,   V_L \psi^{\dagger}_L(0)\psi_L(0)
  ] \psi^{\dagger}_{\nu^{'}}(x^{'},t^{'}) >
  \\&
   + <  [\psi_{\nu}(x,t) ,    V_1 \mbox{  } \psi^{\dagger}_R(0)\psi_L(0) ] \psi^{\dagger}_{\nu^{'}}(x^{'},t^{'}) >
   + <  [\psi_{\nu}(x,t) ,   V^{*}_1 \mbox{  } \psi^{\dagger}_L(0)\psi_R(0)] \psi^{\dagger}_{\nu^{'}}(x^{'},t^{'}) >
\khatamq
or\small
\shuruq
 i{\text{ $v_F$ }}   \mbox{  }\frac{ \Gamma^{\nu,\nu^{'}}(x,x^{'}) }{( \nu x - \nu^{'} x^{'} - {\text{ $v_F$ }} (t-t^{'}) )^2 } \mbox{  } &=  \mbox{ }   i {\text{ $v_F$ }}  \mbox{ }
 \frac{ \Gamma^{\nu,\nu^{'}}(x,x^{'}) }{ (\nu x - \nu^{'} x^{'} - {\text{ $v_F$ }} (t-t^{'}))^2 }\mbox{ }  -i {\text{ $v_F$ }}  \nu\mbox{ }
\frac{ [\partial_x\Gamma^{\nu,\nu^{'}}(x,x^{'})] }{ \nu x - \nu^{'} x^{'} - {\text{ $v_F$ }} (t-t^{'}) }
\\&
+ \delta_{\nu,1} \delta(x) \mbox{  }  V_R  <  \psi_R(0,t)  \psi^{\dagger}_{\nu^{'}}(x^{'},t^{'}) >
   + V_L  \delta_{ \nu,-1} \delta(x) < \psi_L(0,t)
 \psi^{\dagger}_{\nu^{'}}(x^{'},t^{'}) >
  \\&
   + V_1 \delta(x) \delta_{\nu, 1} <  \psi_L(0,t)  \psi^{\dagger}_{\nu^{'}}(x^{'},t^{'}) >
   + V^{*}_1 \delta_{ \nu,-1} \delta(x) \mbox{  }  <  \psi_R(0,t)  \psi^{\dagger}_{\nu^{'}}(x^{'},t^{'}) >
\khatamq\normalsize
or\footnotesize
\shuruq
0 =    &\mbox{ }  -i {\text{ $v_F$ }}  \nu\mbox{ }
\frac{ \sum_{ \gamma,\gamma^{'} = \pm 1} \delta(  x) \gamma \theta(\gamma^{'} x^{'}) \mbox{    }
g_{\gamma,\gamma^{'}}(\nu,\nu^{'})  }{ \nu x - \nu^{'} x^{'} - {\text{ $v_F$ }} (t-t^{'}) }
+ \delta_{\nu,1} \delta(x) \mbox{  }  V_R
\frac{ \Gamma^{1,\nu^{'}}(0,x^{'}) }{  - \nu^{'} x^{'} - {\text{ $v_F$ }} (t-t^{'}) }  + V_L  \delta_{ \nu,-1} \delta(x)
 \frac{ \Gamma^{-1,\nu^{'}}(0,x^{'}) }{   - \nu^{'} x^{'} - {\text{ $v_F$ }} (t-t^{'}) }
  \\&
   + V_1 \delta(x) \delta_{\nu, 1}
   \frac{ \Gamma^{-1,\nu^{'}}(0,x^{'}) }{   - \nu^{'} x^{'} - {\text{ $v_F$ }} (t-t^{'}) }
   + V^{*}_1 \delta_{ \nu,-1} \delta(x) \mbox{  }
   \frac{ \Gamma^{1,\nu^{'}}(0,x^{'}) }{   - \nu^{'} x^{'} - {\text{ $v_F$ }} (t-t^{'}) }
\khatamq\normalsize
or
\shuruq
0 =  &\mbox{ }  -i {\text{ $v_F$ }}  \nu\mbox{ }
\frac{ \sum_{ \gamma,\gamma^{'} = \pm 1} \delta(  x) \gamma \theta(\gamma^{'} x^{'}) \mbox{    }
g_{\gamma,\gamma^{'}}(\nu,\nu^{'})  }{  - \nu^{'} x^{'} - {\text{ $v_F$ }} (t-t^{'}) }
+ \delta_{\nu,1} \delta(x) \mbox{  }  V_R
\frac{
\sum_{ \gamma,\gamma^{'} = \pm 1} \frac{1}{2} \theta(\gamma^{'} x^{'}) \mbox{    }
g_{\gamma,\gamma^{'}}(1,\nu^{'})
 }{  - \nu^{'} x^{'} - {\text{ $v_F$ }} (t-t^{'}) }
 \\ &  + V_L  \delta_{ \nu,-1} \delta(x)
 \frac{ \sum_{ \gamma,\gamma^{'} = \pm 1} \frac{1}{2} \theta(\gamma^{'} x^{'}) \mbox{    }
g_{\gamma,\gamma^{'}}(-1,\nu^{'})  }{   - \nu^{'} x^{'} - {\text{ $v_F$ }} (t-t^{'}) }
   + V_1 \delta(x) \delta_{\nu, 1}
   \frac{
   \sum_{ \gamma,\gamma^{'} = \pm 1} \frac{1}{2} \theta(\gamma^{'} x^{'}) \mbox{    }
g_{\gamma,\gamma^{'}}(-1,\nu^{'})
    }{   - \nu^{'} x^{'} - {\text{ $v_F$ }} (t-t^{'}) }
   \\&+ V^{*}_1 \delta_{ \nu,-1} \delta(x) \mbox{  }
   \frac{\sum_{ \gamma,\gamma^{'} = \pm 1} \frac{1}{2} \theta(\gamma^{'} x^{'}) \mbox{    }
g_{\gamma,\gamma^{'}}(1,\nu^{'})
  }{   - \nu^{'} x^{'} - {\text{ $v_F$ }} (t-t^{'}) }
\khatamq
or
\shuruq
 0 =    &\sum_{ \gamma = \pm 1} (  -i {\text{ $v_F$ }}  \nu\mbox{ }
   \gamma  \mbox{    }
g_{\gamma,\gamma^{'}}(\nu,\nu^{'})
+ \delta_{\nu,1}   \mbox{  }  V_R
  \frac{1}{2}   \mbox{    }
g_{\gamma,\gamma^{'}}(1,\nu^{'})
    + V_L  \delta_{ \nu,-1}
 \frac{1}{2}   \mbox{    }
g_{\gamma,\gamma^{'}}(-1,\nu^{'})
 \\ &  + V_1   \delta_{\nu, 1}
     \frac{1}{2}  \mbox{    }
g_{\gamma,\gamma^{'}}(-1,\nu^{'})
   + V^{*}_1 \delta_{ \nu,-1}   \mbox{  }
    \frac{1}{2}  \mbox{    }
g_{\gamma,\gamma^{'}}(1,\nu^{'}) )
\khatamq
The above equation gives,
\shuru
& V_R = V_L = \frac{2 i {\text{ $v_F$ }} (T-T^*)}{2 T T^*+T+T^*} \\ & V_1 = V_1^{*} =  -\frac{4 i  {\text{ $v_F$ }}\mbox{     } R^* T}{2 T T^*+T+T^*} =
 \frac{4 i {\text{ $v_F$ }}\mbox{     } R T^* }{2 T T^*+T+T^*}
\khatam
This automatically means
\shuru
 -   R^* T  =
 R T^*\\\\
 \khatam


\section*{References}
\bibliographystyle{iopart-num}
\bibliography{ref}

\providecommand{\newblock}{}
\begin{thebibliography}{10}
\expandafter\ifx\csname url\endcsname\relax
  \def\url#1{{\tt #1}}\fi
\expandafter\ifx\csname urlprefix\endcsname\relax\def\urlprefix{URL }\fi
\providecommand{\eprint}[2][]{\url{#2}}

\bibitem{kane1992transport}
Kane C and Fisher M~P 1992 {\em Physical Review Letters\/} {\bf 68} 1220

\bibitem{artemenko2005low}
Artemenko S~N and Remizov S 2005 {\em Physical Review B\/} {\bf 72} 125118

\bibitem{dinh2010tunneling}
Dinh S~N, Bagrets D~A and Mirlin A~D 2010 {\em Physical Review B\/} {\bf 81}
  081306

\bibitem{kainaris2018transmission}
Kainaris N, Carr S~T and Mirlin A~D 2018 {\em Physical Review B\/} {\bf 97}
  115107

\bibitem{lo2019crossover}
Lo C~Y, Fukusumi Y, Oshikawa M, Kao Y~J, Chen P {\em et~al.\/} 2019 {\em
  Physical Review B\/} {\bf 99} 121103

\bibitem{kane1992transmission}
Kane C and Fisher M~P 1992 {\em Physical Review B\/} {\bf 46} 15233

\bibitem{kane1992resonant}
Kane C and Fisher M~P 1992 {\em Physical Review B\/} {\bf 46} 7268

\bibitem{grishin2004functional}
Grishin A, Yurkevich I~V and Lerner I~V 2004 {\em Physical Review B\/} {\bf 69}
  165108

\bibitem{matveev1993tunneling}
Matveev K, Yue D and Glazman L 1993 {\em Physical review letters\/} {\bf 71}
  3351

\bibitem{samokhin1998lifetime}
Samokhin K 1998 {\em Journal of Physics: Condensed Matter\/} {\bf 10} L533

\bibitem{qin1996impurity}
Qin S, Fabrizio M and Yu L 1996 {\em Physical Review B\/} {\bf 54} R9643

\bibitem{hamamoto2008numerical}
Hamamoto Y, Imura K~I and Kato T 2008 {\em Physical Review B\/} {\bf 77} 165402

\bibitem{freyn2011numerical}
Freyn A and Florens S 2011 {\em Physical Review Letters\/} {\bf 107} 017201

\bibitem{ejima2009luttinger}
Ejima S and Fehske H 2009 {\em EPL (Europhysics Letters)\/} {\bf 87} 27001

\bibitem{moon1993resonant}
Moon K, Yi H, Kane C, Girvin S and Fisher M~P 1993 {\em Physical Review
  Letters\/} {\bf 71} 4381

\bibitem{haldane1981luttinger}
Haldane F 1981 {\em Journal of Physics C: Solid State Physics\/} {\bf 14} 2585

\bibitem{giamarchi2004quantum}
Giamarchi T 2004 {\em Quantum Physics in One dimension\/} (Clarendon Oxford)

\bibitem{von1998bosonization}
Von~Delft J and Schoeller H 1998 {\em Annalen der Physik\/} {\bf 7} 225--305

\bibitem{rylands2016quantum}
Rylands C and Andrei N 2016 {\em Physical Review B\/} {\bf 94} 115142

\bibitem{haldane1981demonstration}
Haldane F 1981 {\em Physics Letters A\/} {\bf 81} 153--155

\bibitem{das2018quantum}
Das J~P and Setlur G~S 2018 {\em International Journal of Modern Physics A\/}
  1850174

\bibitem{das2019nonchiral}
Das J~P, Chowdhury C and Setlur G~S 2019 {\em Theoretical and Mathematical
  Physics\/} {\bf 199} 736--760

\bibitem{10.1088/1402-4896/ab957f}
Das J~P, Chowdhury C and Setlur G~S 2020 {\em Physica Scripta\/}

\bibitem{iucci2007fourier}
Iucci A, Fiete G~A and Giamarchi T 2007 {\em Physical Review B\/} {\bf 75}
  205116

\bibitem{schulz1993wigner}
Schulz H 1993 {\em Physical Review Letters\/} {\bf 71} 1864

\bibitem{parola1990asymptotic}
Parola A and Sorella S 1990 {\em Physical Review Letters\/} {\bf 64} 1831

\bibitem{stephan1996dynamical}
Stephan W and Penc K 1996 {\em Physical Review B\/} {\bf 54} R17269

\bibitem{caux2006dynamical}
Caux J~S and Calabrese P 2006 {\em Physical Review A\/} {\bf 74} 031605

\bibitem{gambetta2014correlation}
Gambetta F, Ziani N~T, Cavaliere F and Sassetti M 2014 {\em EPL (Europhysics
  Letters)\/} {\bf 107} 47010

\bibitem{protopopov2011many}
Protopopov I, Gutman D and Mirlin A 2011 {\em Journal of Statistical Mechanics:
  Theory and Experiment\/} {\bf 2011} P11001

\bibitem{sen1999density}
Sen D and Bhaduri R 1999 {\em Canadian Journal of Physics\/} {\bf 77} 327--341

\bibitem{aristov2007luttinger}
Aristov D 2007 {\em Physical Review B\/} {\bf 76} 085327

\bibitem{PhysRevLett.69.2863}
White S~R 1992 {\em Phys. Rev. Lett.\/} {\bf 69}(19) 2863--2866

\bibitem{RevModPhys.77.259}
Schollw\"ock U 2005 {\em Rev. Mod. Phys.\/} {\bf 77}(1) 259--315

\bibitem{Schneider2006ConductanceIS}
Schneider G and Schmitteckert P 2006 Conductance in strongly correlated 1d
  systems: Real-time dynamics in dmrg

\bibitem{PhysRevLett.113.070601}
Kantian A, Schollw\"ock U and Giamarchi T 2014 {\em Phys. Rev. Lett.\/} {\bf
  113}(7) 070601

\bibitem{Ejima_2009}
Ejima S and Fehske H 2009 {\em {EPL} (Europhysics Letters)\/} {\bf 87} 27001

\bibitem{PhysRevB.56.9766}
Qin S, Fabrizio M, Yu L, Oshikawa M and Affleck I 1997 {\em Phys. Rev. B\/}
  {\bf 56}(15) 9766--9774

\bibitem{PhysRevB.99.121103}
Lo C~Y, Fukusumi Y, Oshikawa M, Kao Y~J and Chen P 2019 {\em Phys. Rev. B\/}
  {\bf 99}(12) 121103

\bibitem{PhysRevLett.116.247204}
Khemani V, Pollmann F and Sondhi S~L 2016 {\em Phys. Rev. Lett.\/} {\bf
  116}(24) 247204

\bibitem{PhysRevLett.118.017201}
Yu X, Pekker D and Clark B~K 2017 {\em Phys. Rev. Lett.\/} {\bf 118}(1) 017201

\bibitem{Oshi}
Lo C~Y, Fukusumi Y, Oshikawa M, Kao Y~J and Chen P 2019 {\em Phys. Rev. B\/}
  {\bf 99}(12) 121103

\bibitem{das2017one}
Das J~P and Setlur G~S 2017 {\em Physica E: Low-dimensional Systems and
  Nanostructures\/} {\bf 94} 216--230

\bibitem{das2018ponderous}
Das J~P and Setlur G~S 2018 {\em EPL (Europhysics Letters)\/} {\bf 123} 27002

\bibitem{DAS201939}
Das J~P and Setlur G~S 2019 {\em Physica E: Low-dimensional Systems and
  Nanostructures\/} {\bf 110} 39 -- 48 ISSN 1386-9477

\bibitem{Das2019C}
Das J~P and Setlur G~S 2019 {\em Physics Letters A\/} {\bf 383} 3149 -- 3161
  ISSN 0375-9601

\end{thebibliography}
\end{document}